\documentclass[10pt]{iopart} 
\usepackage[english]{babel}
\usepackage{iopams} 

\expandafter\let\csname equation*\endcsname\relax
\expandafter\let\csname endequation*\endcsname\relax
\usepackage{amsmath}
\usepackage{amssymb}
\usepackage{color}
\usepackage{geometry}
\usepackage{ifthen} 
\usepackage[pdftex]{graphicx}
\usepackage[pdftex]{hyperref} 

\hypersetup{colorlinks=true,
	linkcolor=blue,
	citecolor=blue,
	urlcolor=blue,
	pdfauthor={Paul Heinrich IPP, paul.heinrich@ipp.mpg.de},
	pdfsubject={Heinrich 2026 Evolution of SPI-induced disruptions at ASDEX Upgrade},
	pdftitle={Evolution of SPI-induced disruptions in ASDEX Upgrade}
}

\DeclareGraphicsExtensions{.pdf}
\graphicspath{{graphics/}}

\usepackage[numbers,sort&compress]{natbib}
\usepackage{hypernat}

\usepackage{bmpsize}
\usepackage{booktabs}
\usepackage{dblfloatfix}
\usepackage{caption}
\usepackage[section]{placeins}
\usepackage{xspace}
\usepackage{mwe}
\usepackage[normalem]{ulem}
\usepackage{makecell}

\usepackage{xargs}
\usepackage[pdftex,dvipsnames]{xcolor}  % Coloured text etc.
\usepackage[colorinlistoftodos,prependcaption,textsize=tiny]{todonotes}
\newcommandx{\unsure}[2][1=]{\todo[linecolor=red,backgroundcolor=red!25,bordercolor=red,#1]{#2}{\color{red}\textbf{?}}}
\newcommandx{\change}[2][1=]{\todo[linecolor=blue,backgroundcolor=blue!25,bordercolor=blue,#1]{#2}{\color{blue}\textbf{!}}}
\newcommandx{\toadd}[2][1=]{\todo[linecolor=cyan,backgroundcolor=cyan!25,bordercolor=cyan,#1]{#2}{\color{cyan}\textbf{\ensuremath{\forall}}}}
\newcommandx{\improvement}[2][1=]{\todo[linecolor=Plum,backgroundcolor=Plum!25,bordercolor=Plum,#1]{#2}}
\newcommandx{\ask}[3][1=]{\todo[linecolor=OliveGreen,backgroundcolor=OliveGreen!25,bordercolor=OliveGreen,#1]{@#2:\\#3}{\color{OliveGreen}\textbf{?}}}
            
\definecolor{pythonblue}{RGB}{31,119,180}
\definecolor{pythonorange}{RGB}{255,127,14}
\definecolor{pythongreen}{RGB}{44,160,44}
\definecolor{pythoncoolwarmblue}{RGB}{117,151,246}
\definecolor{pythoncoolwarmbrown}{RGB}{237,209,194}
\definecolor{pythoncoolwarmorange}{RGB}{242,146,116}

\newcommand{\etalc}[1]{\etal~\cite{#1}}

\usepackage[export]{adjustbox}

\newcommand{\fig}[1]{figure~\ref{fig:#1}\xspace}
\newcommand{\subfig}[2]{\mbox{figure~\ref{fig:#1}(#2)}\xspace}

\newcommand{\sect}[1]{section~\ref{sec:#1}\xspace}
\newcommand{\tabl}[1]{table~\ref{tab:#1}\xspace}

\newcommand{\dtCQ}[2]{\ensuremath{\Delta \textrm{t}_\textrm{CQ}^{#1 \rightarrow #2}}}

\begin{document}
\hyphenation{ASDEX}

\title{Evolution of SPI-induced disruptions in ASDEX Upgrade}
\author{P.~Heinrich$^1$\footnote[7]{corresponding author: paul.heinrich@ipp.mpg.de}, G.~Papp$^1$, S.~Jachmich$^2$, J.~Artola$^2$, M.~Bernert$^1$, P.~de~Marn\'e$^1$, M.~Dibon$^2$, R.~Dux$^1$, T.~Eberl$^1$, O.~Ficker$^3$, P.~Halldestam$^1$, J.~Hobirk$^1$, M.~Hoelzl$^1$, F.~Klossek$^1$, M.~Lehnen$^2$\footnote[2]{Deceased}, T.~Lunt$^1$, M.~Maraschek$^1$, A.~Patel$^1$, T.~Peherstorfer$^4$, N.~Schwarz$^5$, U.~Sheikh$^6$, B.~Sieglin$^1$, J.~Svoboda$^3$, W.~Tang$^1$, the ASDEX Upgrade Team$^a$ and the EUROfusion Tokamak Exploitation Team$^{b}$}
\address{$^1$Max Planck Institute for Plasma Physics, Boltzmannstr. 2, D-85748 Garching, Germany}
\address{$^2$ITER Organization, Route de Vinon-sur-Verdon, CS 90 046 13067 St.~Paul-lez-Durance, France}
\address{$^3$Institute of Plasma Physics of the CAS, CZ-18200 Praha 8, Czech Republic}
\address{$^4$Institute for Applied Physics, Wiedner Hauptstr. 8-10/134, 1040 Wien, Austria}
\address{$^5$Commissariat \`{a} l'\'{E}nergie Atomique (CEA), Institute for Magnetic Fusion Research (IRFM), F-13108 St.~Paul-lez-Durance, France}
\address{$^6$\'{E}cole Polytechnique F\'{e}d\'{e}rale de Lausanne (EPFL), Swiss Plasma Center (SPC), CH-1015 Lausanne, Switzerland}
\address{$^a$See the author list of \href{https://doi.org/10.1088/1741-4326/ad249d}{H.~Zohm~\etal 2024 Nucl. Fusion}} 
\address{$^b$See the author list of \href{https://doi.org/10.1088/1741-4326/ad2be4}{E.~Joffrin~\etal 2024 Nucl. Fusion}}

\date{\today}

\begin{abstract}
    Disruptions are a major concern for future fusion reactors based on the tokamak principle. To ensure machine protection, the thermal loads and vessel forces that arise during disruptions have to be mitigated reliably. For the ITER disruption mitigation system~(DMS), the shattered pellet injection~(SPI) technology has been selected. It can provide a prompt delivery of the injection material into the plasma core, with the mitigation efficiency depending on fragment size and velocity. A highly flexible SPI system was built and installed at the tokamak ASDEX Upgrade~(AUG) to aid the finalization process of the ITER DMS and provide crucial input for modeling. The SPI-induced disruptions in the 2022 AUG experiments follow a typical chain of events, which are discussed in this paper: The first light~(FL), main fragment arrival~(MFA), plasma movement event~(PME), MARFE, thermal quench~(TQ)/plasma current spike~(IP-spike), current quench~(CQ), and vertical displacement event~(VDE) phase.
    Depending on the injection parameters, these phases may vary significantly or some might not be present at all.
    In this paper, we will focus on the characterization of these disruption phases and figures of merit for the mitigation efficiency, depending on the SPI configuration.
    With increasing amount of assimilated neon in the plasma -- primarily influenced by the neon content in the pellet but also the shattering parameters -- the disruptions exhibit different behaviors. This disruption evolution seems to be a continuous process, with the most prominent feature being the changing disruption time scales and plasma current time trace shape during the CQ from convex (poorly or unmitigated) $\rightarrow$ concave (well mitigated/radiation dominated). Depending on the injection, pre-TQ durations between 15 -- 0.5~ms and early CQ durations ($\Delta \textrm{t}_\textrm{CQ}^{100 \rightarrow 80}$) between 13.3 -- 8.2~ms had been achieved at AUG.
\end{abstract}

\submitto{\NF}

\maketitle
\ioptwocol 

\section{Introduction \label{sec:Introduction}}

Disruptions pose a significant threat to machine integrity and operation of high-current fusion experiments and future reactors of the tokamak type~\cite{Noll_1989, Noll_1997, Gerasimov_2020, Artola_2024, Hollmann_2011_PFCdisruptions, Arnoux_2011, Lehnen_2013b}.
In a disruption the plasma current, necessary to provide a stable equilibrium confinement of the plasma, can terminate and with it the energy stored in the plasma is released~\cite{Gorbunov_1963, Hender_2007_ITER_Physics, Hollmann_2011_PFCdisruptions, Boozer_2012, ITER_1999_Physics_Basis}. If not mitigated, high heat loads~\cite{Hollmann_2011_PFCdisruptions, Arnoux_2011, Lehnen_2013b} and electromagnetic forces~\cite{Noll_1989, Noll_1997, Gerasimov_2020, Artola_2024} will arise. The toroidal electric field induced in the decaying plasma can even create so-called runaway electron~(RE) beams~\cite{Dreicer_1959, Dreicer_1960, Gurevich_1961, Connor_1975, Breizman_2019} that can cause strongly localised heat loads at impact with the plasma facing components~\cite{Nygren_1997, Reux_2015, Matthews_2016, Bandaru_2024, Bergstroem_2024}. If a disruption can no longer be avoided~\cite{Sieglin_2024_MARFE, Kudlacek_2024_DCS, deVries_2024}, robust disruption mitigation action has to be taken. In present day machines disruption mitigation typically involves the injection of large quantities of material into the plasma, which is usually several times the plasma inventory~\cite{Hender_2007_ITER_Physics, Pautasso_2015, Lehnen_ITER_workshop_2021}. The injected material is typically neon~(Ne) or argon~(Ar) (or gas mixtures containing either of the two) to spread the heat loads over a large surface area via radiation. Pure neon massive gas injection~(MGI) is routinely used to mitigate disruptions in many present day devices, such as ASDEX Upgrade~(AUG)~\cite{Treutterer_2014, Dibon_2017_PhD, Pautasso_2020_MGI_exp}. In order to increase the penetration and deposition depth, and to improve system response time, the shattered pellet injection~(SPI) technique was first tested at DIII-D in 2009~\cite{Commaux_2010, Baylor_2019_SPI-DIII-D-JET}, and will be used for the ITER disruption mitigation system~(DMS)~\cite{Lehnen_ITER_workshop_2021,Lehnen_2023_FEC, Loarte_2024}.
For ITER, a cryogenic pellet of protium~($\textrm{H}_2$), Ne or a $\textrm{H}_2$-Ne-mixture is accelerated via a propellant gas towards the plasma. Before it leaves the guide tube~(GT), it will shatter inside the shattering unit close to the plasma edge with the resulting fragments entering the plasma to increase the surface-to-volume ratio for rapid ablation and assimilation of the material~\cite{Baylor_2010_DMS_SPI}.

To aid the design process, provide crucial input for the ITER DMS, and support modelling efforts, a highly flexible SPI system was deployed at AUG~\cite{Dibon_2023_SPI}. This project was accompanied by a large diagnostic upgrade to the tokamak, including foil bolometer measurements with matching viewing geometry in 5 toroidally separated sectors~(S), additional Absolute eXtended UltraViolet~(AXUV) diodes and ultra-highspeed~(UHS) cameras viewing from the side and the top onto the injection. This setup allows for a detailed analysis of the radiation characteristics~\cite{Heinrich_2025_frad, Heinrich_2025_PhD}. Prior to the plasma experiments in AUG, extensive laboratory experiments were conducted~\cite{Dibon_2023_SPI, Heinrich_2024_SPI_Lab, Kovacs_2024, Heinrich_2025_PhD}, which investigated the shattering process of the pellets to determine the resulting fragment size and velocity distributions~\cite{Peherstorfer_2022_fragmentation, Illerhaus_2024_SPI, Lee_2024_PD, Lee_2024_PhD}. This characterization of these distributions from the laboratory experiments enables to investigate their influence on the tokamak disruptions -- i.e. material assimilation~\cite{Jachmich_2023_EPS, Jachmich_2024_EPS} and radiation characteristics~\cite{Heinrich_2025_PhD, Heinrich_2025_frad}, 
used as indicators for different disruption phases and their evolution.

In this paper, we analyse these dynamics and the characteristic phases of the SPI-induced disruptions and their evolution with changing pellet and shattering parameters for the 2022 experiments at AUG.
This set of experiments also included a wide range of pellets with different percentage of neon in the pellet~($\text{f}_\textrm{neon}$) to study the effect of plasmoid drift suppression on the material assimilation~\cite{Vallhagen_2025}. While for the thermal load mitigation high neon fractions are relevant, the assimilation of deuterium~(D2) for RE-beam suppression was studied using pure~$\textrm{D}_2$ and slightly neon doped pellets.

The paper is structured as follows: in \sect{Experimental_setup}, the experimental setup relevant for this paper is introduced. Subsequently, the individual disruption phases are illustrated in \sect{Phases} and their evolution in \sect{Evolution}. Finally, the disruption timescales are provided in \sect{Time_scales}.

\section{Experimental setup \label{sec:Experimental_setup}}

\begin{figure}[htb]
	\centering
	\includegraphics[width=\linewidth]{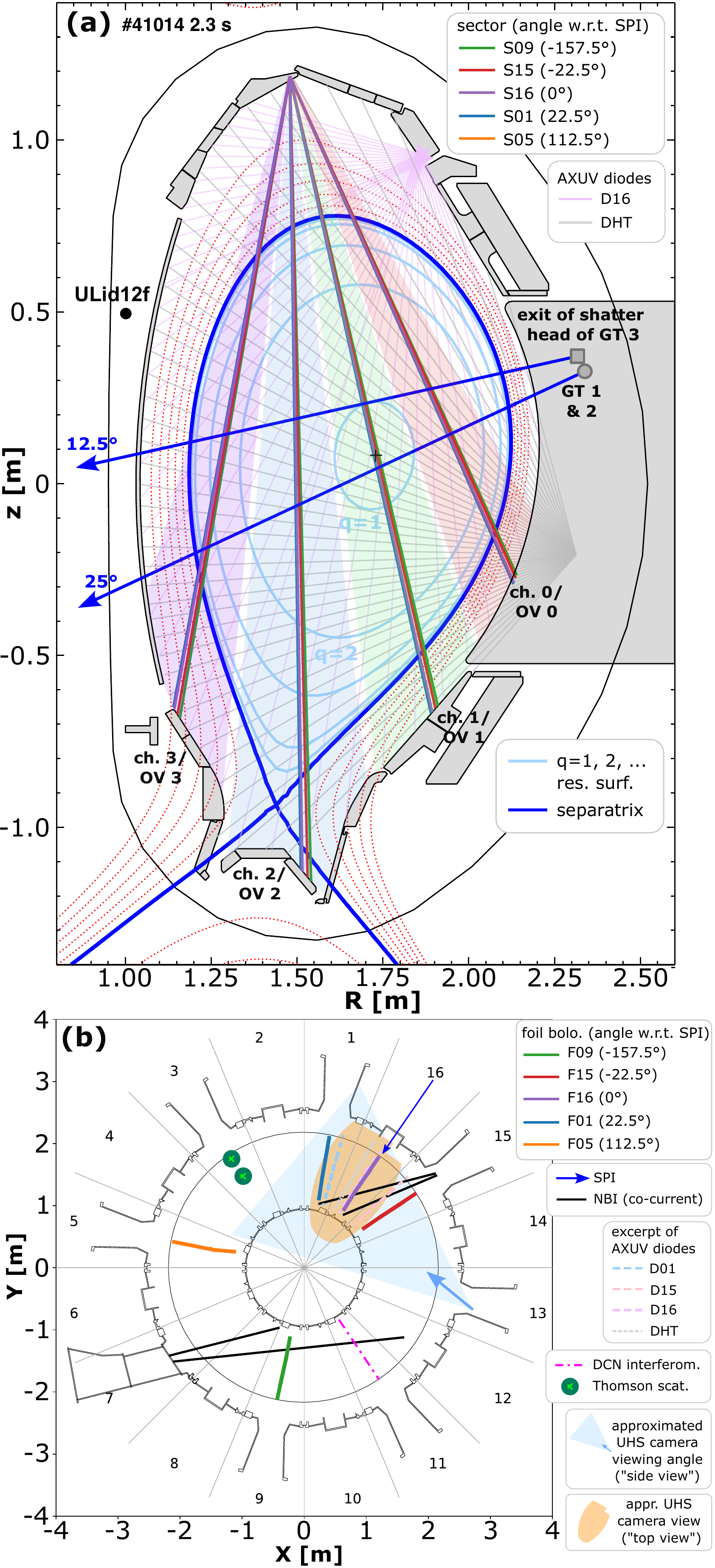}
	\caption{Poloidal~(a) and toroidal~(b) cross section of the AUG vacuum vessel indicating diagnostic measurement locations and the injection geometry. Four distinct observation volumes are created for each channel of the foil bolometers (full-shadow volume depicted) and the central lines are only added for illustration of the viewing angles also in~(b). \label{fig:exp_setup_foils}}
\end{figure}

The AUG SPI has three independent injection barrels, with three independent cold cells (for pellet desublimation), individual GTs, and separate shatter heads which are installed at the end of each GT. The three different shatter heads attached for the 2022 experiments allowed to disentangle the effects of fragment size and velocity distributions on the disruptions. The independent GTs also allow for simultaneous or synchronised injections. For the experiments in 2022, the following three shatter heads were installed~\cite{Dibon_2023_SPI, Heinrich_2024_SPI_Lab, Heinrich_2025_PhD, Heinrich_2025_frad}:

\begin{enumerate}
	\item[GT1] \textbf{$25^{\circ}$, rectangular cross-section, long:} producing a collimated plume of smaller \& slower fragments. By matching the perpendicular velocity component~($\text{v}_\perp$) component w.r.t. the $12.5^{\circ}$ head, similar fragment size distributions are expected but with lower mean fragment velocity~($\text{v}_\textrm{fragment}$). This way, fragment size \& velocity effects can be disentangled.
	\item[GT2] \textbf{$25^{\circ}$, circular cross-section, short:}
	producing a fragment plume with increased spatial spread of smaller \& slower fragments. Note, the effective shattering angle depends on the exact impact position of the pellet in the circular head, hence the fragmentation process shows a larger variance in fragment size \& velocity distributions and fragments can show a swirling motion due to rotation inside the head~\cite{Peherstorfer_2022_fragmentation, Heinrich_2025_frad}.
	\item[GT3] \textbf{$12.5^{\circ}$, rectangular cross-section, long:} this head allows to inject a collimated plume of larger \& faster fragments for the same pre-shattering pellet velocity~($\text{v}_\textrm{pellet}$) compared to heads with a larger shattering angle.
\end{enumerate}

\begin{figure*}[htb]
	\centering
	\includegraphics[width=\textwidth]{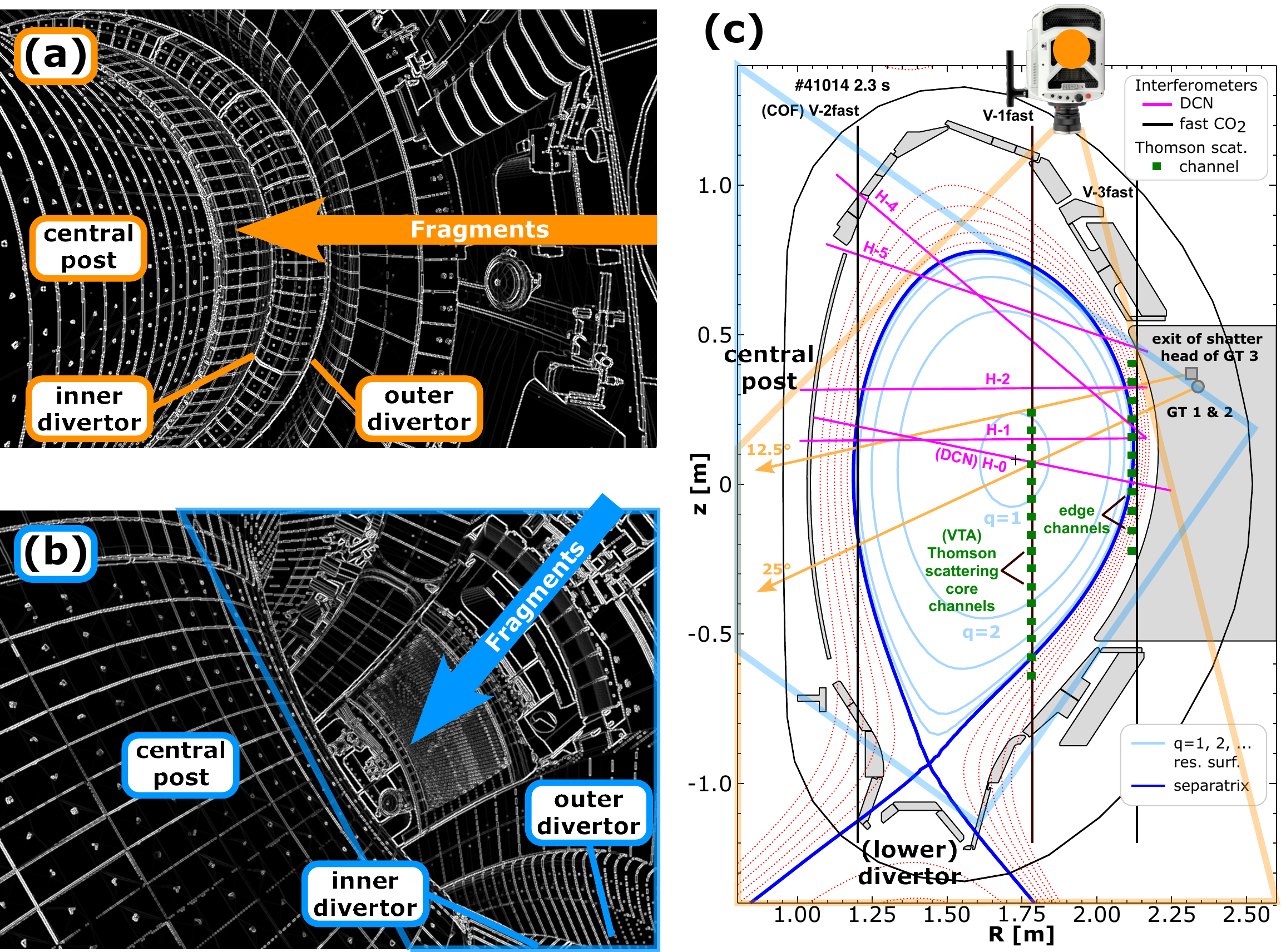}
	\caption{Virtual reconstruction of the top-down~(a) and toroidal~(b) views of the UHS cameras used to analyse the fragment ablation and disruption processes. The orange/light blue frames in~(c) represent the approximated viewing geometries from~(a),~(b) in the poloidal cross section, respectively. The cropped version of the image in~(b) (and~(c)) is used in \fig{disruption_phases}. In~(c), the LOSs of the density measurements of the DCN~\cite{Mlynek_2014_Fringe-jumps, Mlynek_2010_DCN} and $\textrm{CO}_2$~\cite{Mlynek_2012_CO2} laser interferometry systems as well as the Thomson scattering channels~\cite{Kurzan_2011} are shown. \label{fig:exp_setup}}
\end{figure*}

Figures~\ref{fig:exp_setup_foils} and~\ref{fig:exp_setup} show an excerpt of diagnostics that are relevant for the discussion on the SPI-induced disruption phases and their evolution. In the poloidal cross-section in \subfig{exp_setup_foils}{a}, the shatter head location \& injection angles, the 4-channel foil bolometers, and an excerpt of the LOSs of the AXUV diodes are shown. Via a pinhole, the four observation volumes~(OVs) indicated for each channel are created and the central line is only added for illustration purposes (not a line-integrated measurement). Figure~\ref{fig:exp_setup_foils}(b) shows the toroidal cross section, which indicates the toroidal distribution of the diagnostic measurements and an estimated viewing cone of the toroidal UHS camera view from \subfig{exp_setup}{b,~c}.

A virtual view of the two UHS cameras is provided in \fig{exp_setup} -- top-down view in~(a) and toroidal view in~(b).
An estimated projection of the view into the poloidal plane is illustrated in~(c) via the orange/light-blue \mbox{tri-/rectangles}, respectively. In addition, the LOSs of the laser interferometry systems and the locations of the Thomson scattering channels are provided in the poloidal cut in~(c).

\begin{figure*}[htb]
	\centering
	\includegraphics[width=\textwidth]{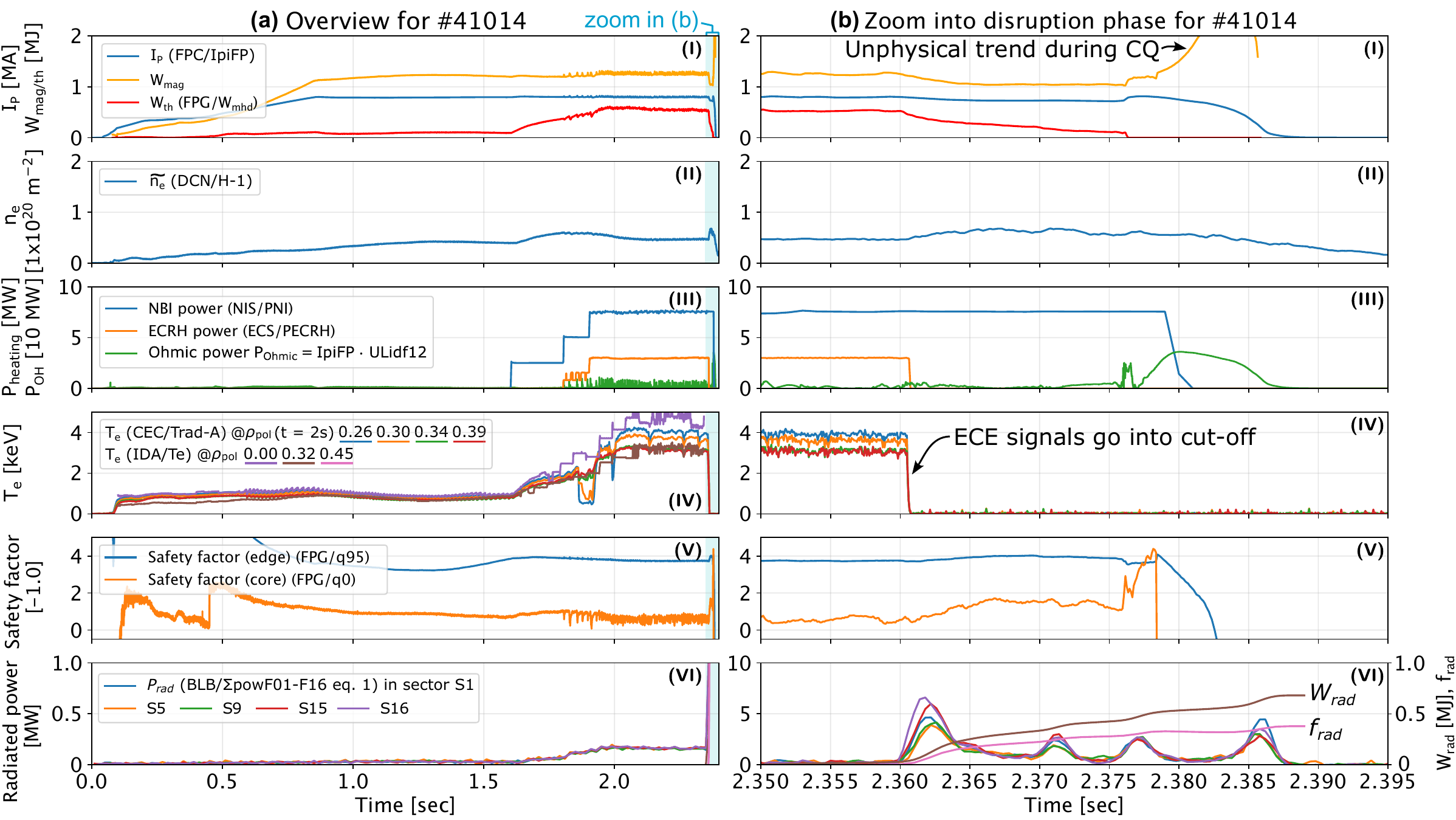}
	\caption{Scenario overview for the reference discharge \#41014 of the SPI H-mode: 100\%~$\textrm{D}_2$ injection via the $25^{\circ}$, rectangular shatter head with $\text{v}_\textrm{pellet} = 466$~m/s. The entire discharge is shown in~(a) and a zoom into the disruption is provided in~(b). The following signals are shown:
	(I)~the plasma current, stored magnetic and thermal energies,
	(II)~line integrated electron density derived from the core LOS H-1~\cite{Mlynek_2010_DCN, Mlynek_2014_Fringe-jumps} (no valid $\textrm{CO}_2$ laser measurement available for this discharge),
	(III)~heating powers ($P_\textrm{NBI}$, $P_\textrm{ECRH}$, $P_\textrm{Ohmic}$),
	(IV)~electron temperature from electron cyclotron emission~\cite{Denk_2018_AUG-ECE} and the integrated data analysis~(IDA)~\cite{Fischer_2010_IDA} are provided. The electron cyclotron emission and IDA signals in~(a) are averaged with $\textrm{w}_\textrm{avg} = 3.2$~ms and 5~ms, respectively. Note, the electron cyclotron emission signals go into density cut-off~\cite{Denk_2018_AUG-ECE} around $\approx 2.36$~sec, hence do not show the thermal quench~(TQ) (compare $W_\textrm{th}$ from~(I)).
	In~(V) the core and edge safety factors ($q$, $q_{95}$) and in~(VI) the $P_\textrm{rad, sector}$-values for the five measurement sectors ($\textrm{w}_\textrm{avg} = 20$~ms in~(a)) are provided. In~(VIb) also the radiated energy~($\textrm{W}_\textrm{rad}$) and radiated energy fraction~($\text{f}_\text{rad}$) are shown. Further details on the calculation of $P_\textrm{rad, sector}$, $\text{W}_\text{rad}$, and $\text{f}_\text{rad}$ are given in the paper by P.~Heinrich~\etal~\cite{Heinrich_2025_frad}. For this 100\%~$\textrm{D}_2$, 8~mm injection, four radiation peaks are visible during the main fragment arrival~(MFA), plasma movement event~(PME), TQ/IP-spike, and VDE phase.\label{fig:scenario}}
\end{figure*}

In this paper, we will focus on single pellet SPI into 1.8~T, 800~kA, H-modes (``SPI H-mode''~\cite{Heinrich_2025_frad, Heinrich_2025_PhD}) for our description of the disruption phases and evolution.
As a reference discharge for this analysis, we selected \#41014, which showed one of the longest disruption durations: in total $\approx 30$~ms with a pre-TQ duration $\Delta \textrm{t}_\textrm{pre-TQ} \gtrsim 15$~ms (see section~\ref{sec:Time_scales}). This allows to also resolve the different radiation peaks -- connected to the disruption phases -- in the foil bolometer signals with a time resolution of 0.8~ms. An overview of the discharge parameters for \#41014 is provided in \fig{scenario}.
In this discharge we injected a 100\%~$\textrm{D}_2$ pellet with $\text{v}_\textrm{pellet} = 466$ m/s via the $25^{\circ}$, rectangular shatter head, producing a lot of tiny fragments and dust~\cite{Heinrich_2025_frad, Heinrich_2025_PhD, Peherstorfer_2022_fragmentation}.
More details on the SPI system, target plasmas, installed shatter heads for the 2022 experiments, and experimental setup (including the calculation of the $\text{W}_\text{rad}$ \& $\text{f}_\text{rad}$) can be found in the paper by P.~Heinrich~\etalc{Heinrich_2025_frad} and his thesis~\cite{Heinrich_2025_PhD}.

\clearpage

\begin{figure*}[b]
	\centering
	\includegraphics[width=\textwidth]{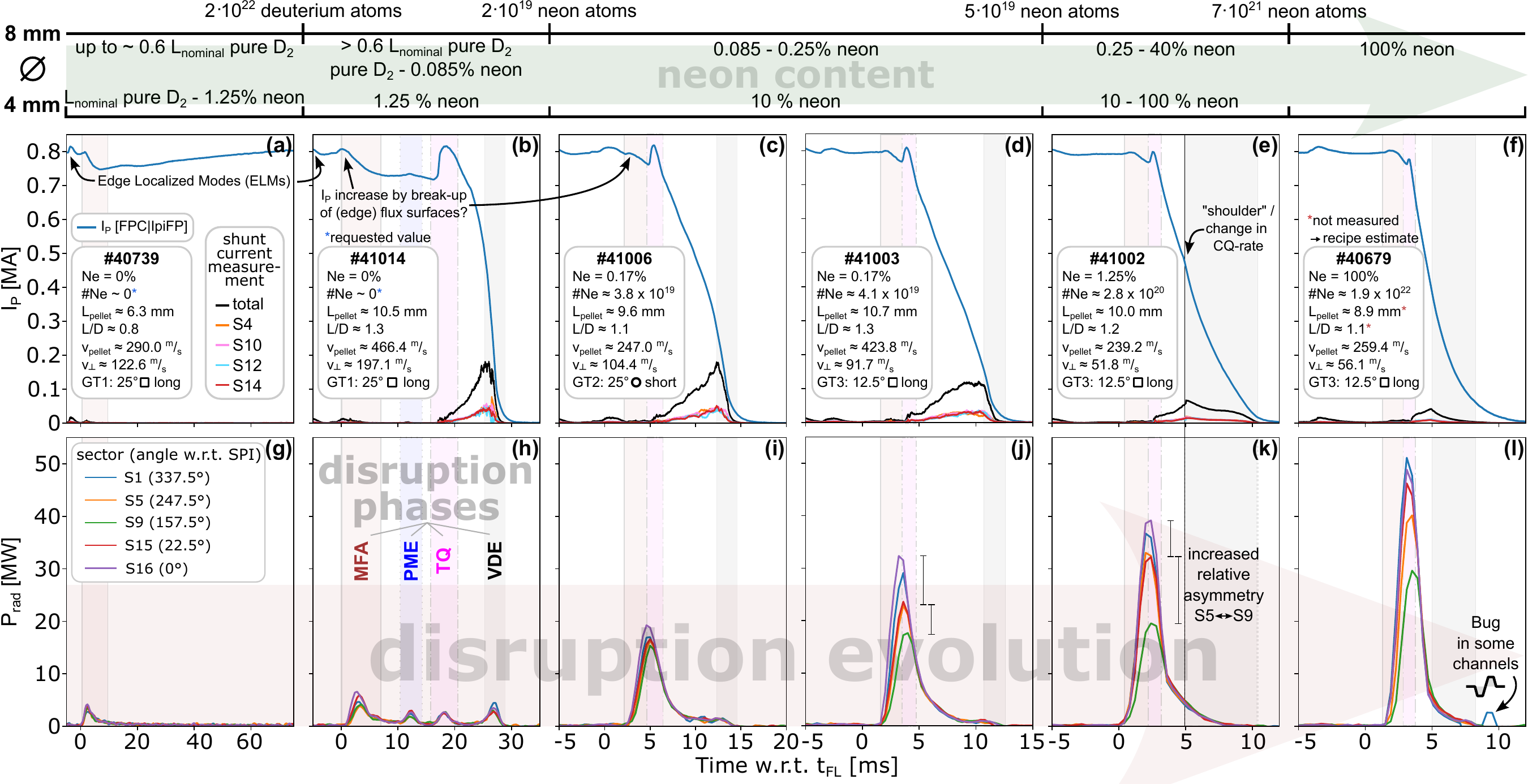}
	\caption{Overview over the evolution of disruption behaviors from 100\%~$\textrm{D}_2$ to 100\%~Ne~SPI. In (a)~$\rightarrow$~(f) the plasma \& shunt current measurements are shown and the radiated power in an individual sector in (g)~$\rightarrow$~(l).
	The individual disruption phases within a discharge -- indicated by the coloured areas (see~(h)) -- are introduced in \sect{Phases}, whereas the evolution (a)~$\rightarrow$~(f) with changing neon content is discussed in \sect{Evolution}. The neon content (atoms or $\text{f}_\textrm{neon}$) is given for the 8~mm~(top) and 4~mm~(bottom) pellets at the top of the image with the tick-markers corresponding to the individual behaviors of the evolution. \label{fig:disruption_evolution_1}}
\end{figure*}

\section{Disruption phases \label{sec:Phases}}
One of the main objectives of the SPI experiments at ASDEX Upgrade is to find the optimal fragment size and velocity distributions -- mainly determined by the shatter head geometry --  and neon concentration for the disruption mitigation. We observed a continuous evolution of the SPI-induced disruptions from conduction dominated (unmitigated) to radiation dominated (well mitigated) disruptions by scanning the neon content in the pellet or changing the fragment sizes and velocities. This disruption evolution is illustrated in~\fig{disruption_evolution_1} and is apparent due to changes in the chain of events during an SPI-induced disruption --~the disruption phases~-- as well as changing time scales between these disruption phases. In the following, we introduce the disruption phases in chronological order using the reference discharge~\#41014 (\subfig{disruption_evolution_1}{b, h}, \fig{disruption_phases}) as example. Afterwards we will discuss their evolution (a)~$\rightarrow$~(f) with changing injection parameters in \sect{Evolution}.

\begin{figure*}[htb]
	\centering
	\includegraphics[width=\textwidth]{figure_5.pdf}	\caption{Images taken from the fast camera recordings of the disruption during the different disruption phases (1--24) with a transparent$\rightarrow$blue$\rightarrow$red intensity colour-mapping (transparency increasing below 33\%, fully transparent below 30\%; compare \fig{MFA}). The virtual view from \subfig{exp_setup}{b} is provided in the background. In~(a) the plasma and shunt currents, and in~(b) the radiation of the individual channels \mbox{ch0 -- ch3} (see figure~\ref{fig:exp_setup_foils}(a)) of the foil bolometer in \mbox{S16} are shown. The regions of highest intensity for each phase are displayed in~(c--e): the MFA and post injection phase in~(c), the PME and MARFE in~(d), and the VDE motion in~(e). A video is provided on the journal webpage.\label{fig:disruption_phases}}
\end{figure*}

\subsection{First Light (FL) \label{sec:FL}}
\begin{figure*}[htb]
	\centering
	\includegraphics[width=\textwidth]{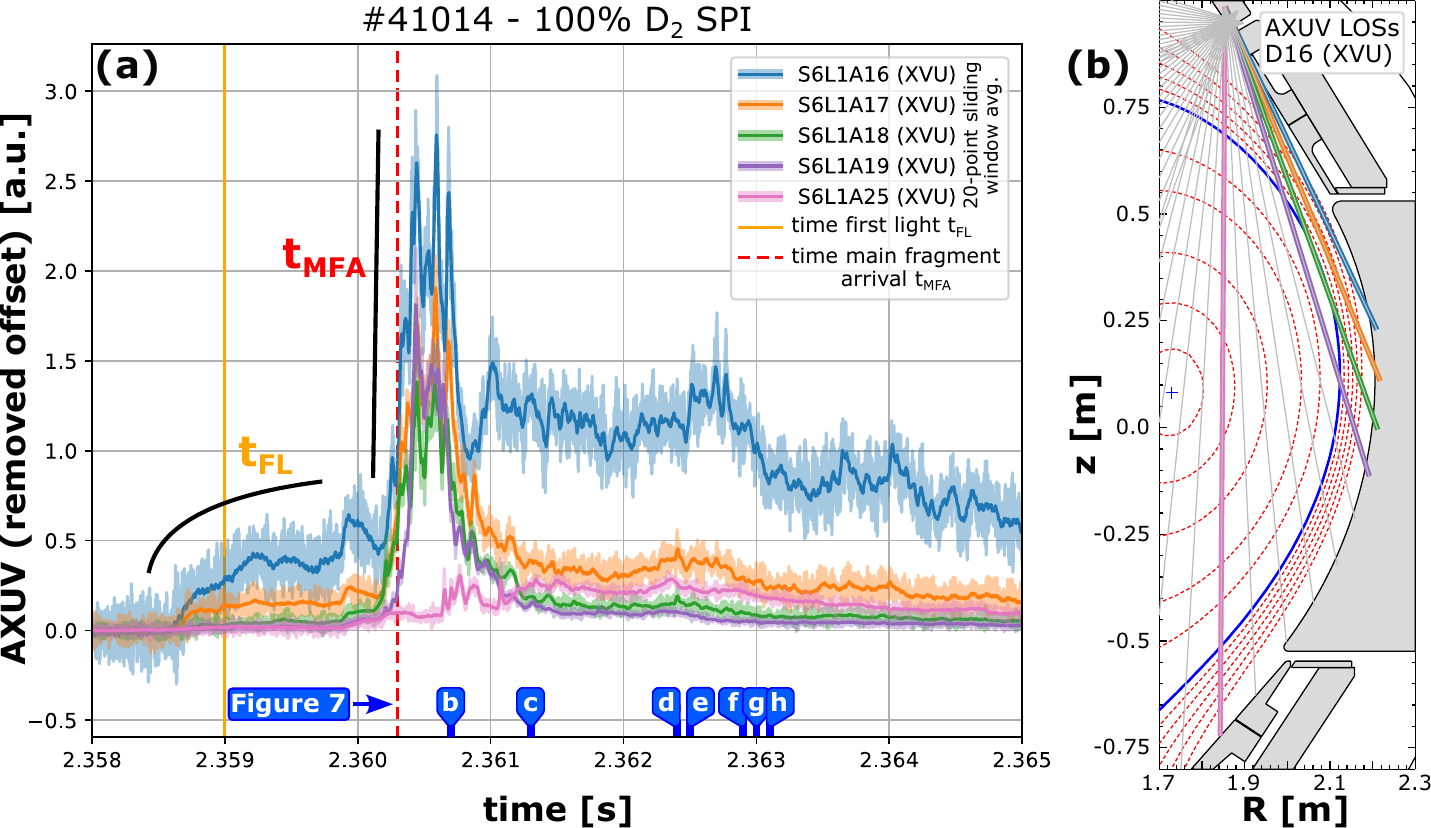}
	\caption{In~(a), the AXUV signals with a 20-point average in dark is given. The first increase in the edge AXUV channels to a plateau-like level is used as the first light~(FL) time marker. Soon afterwards the steep increase in radiation marks the main fragment arrival~(MFA). The LOSs from~(a) are shown in~(b). \label{fig:FL_and_MFA}}
\end{figure*}

Prior to the main fragments entering the plasma, a slight increase in the outermost lines of sight~(LOSs) (directed towards the low field side~(LFS) midplane) of the fast AXUV camera in sector 16 is observed. This small increase to a plateau-like level is referred to as first light~(FL) (see \fig{FL_and_MFA}) and is considered to be caused by first tiny fragments and small amounts of gas entering the chamber followed by the rest of the fragments shortly afterwards. As the AUG~SPI system removes at least 99.96\% of the propellant gas before the pellet reaching the shatter head~\cite{Dibon_2023_SPI, Heinrich_2024_SPI_Lab}, this gas is generated during the shattering process. This first increase in radiation is frequently used as a first time marker --~time of the FL~($\textrm{t}_\textrm{FL}$)~-- not only in AUG, but also in other SPI experiments, such as e.g in Joint European Torus~(JET)~\cite{Sheikh_2024_APS}.

\subsection{Main Fragment Arrival (MFA) \label{sec:MFA}}
As the main bulk of fragments starts to enter the plasma (time of the MFA~($\textrm{t}_\textrm{MFA}$)) a first radiation peak is visible -- which is usually the largest radiation peak (compare figures~\ref{fig:disruption_evolution_1}(a) and~\ref{fig:disruption_evolution_2}(a, e)) throughout the injection/disruption phase. This radiation peak will be also referred to as ``main fragment arrival~(MFA) (radiation) peak''.
As the fragments are injected in sector~16, most of the time the measured radiation is highest in this sector (see \subfig{disruption_evolution_1}{g--k}, violet line). This is especially the case for 100\%~$\textrm{D}_2$ injections (or neon doped pellets with $< 1$\%~Ne), in which case the radiation is dominated by the foil bolometer channel~0 -- directed to the LFS from which the fragments enter the plasma.

\begin{figure*}[htb]
	\centering
	\includegraphics[width=\textwidth]{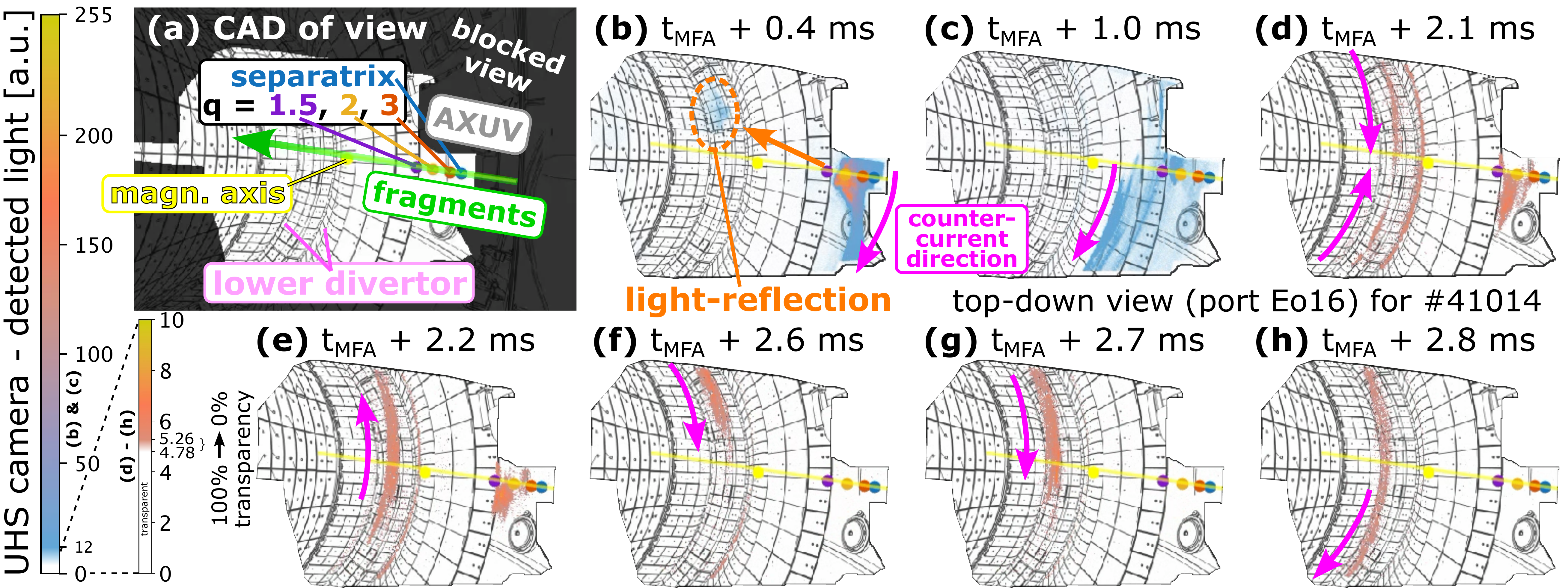}
	\caption{Top-down view on the main fragment arrival with a fast camera for discharge \#41014. The colour-inverted virtual view from \subfig{exp_setup}{a} is provided in the background and the UHS camera recording with transparent$\rightarrow$blue$\rightarrow$yellow intensity colour-mapping on top (transparency increasing for values below 12/5.26, fully transparent below 3.82/4.78).
	The view is partially blocked by components (i.e. tiles, AXUV-box) indicated by the black shaded area in~(a). Only the visible area is cut out in~(b)--(h).
	The fragments arrive at the LFS (right side) and travel towards the plasma center (towards the left). Note, the contrast in the images (d)--(h) is increased to increase visibility of the radiation pattern above the divertor (otherwise too faint). A video is provided on the journal webpage. \label{fig:MFA}}
\end{figure*}

During the 2022 experiments, a fast camera viewing top-down above the injection was observing the fragment arrival in the plasma to study the fragment ablation process, as shown in figures~\ref{fig:exp_setup} and~\ref{fig:MFA}. The radiation pattern spreads around the torus (b~and~c), following the magnetic field lines. The preferred rotation direction seems to depend on the neon content and in this 100\% $\textrm{D}_2$ case, appears to prefer the counter~(ctr)-current direction~\cite{Heinrich_2025_PhD}. Note, while the radiation pattern is most likely directly connected to the location of the ablated material, in general, the radiation is also affected by the local temperature and electron density. However, as the parallel transport timescales are fast compared to the camera time resolution, we assume this radiation pattern to indicate the position of the ablated pellet material.
Around 4~ms after the MFA, a radiation pattern appears in the divertor region, visible in both fast camera recordings: the top-down view (\subfig{MFA}{d}) and the side view (green circle in \subfig{disruption_phases}{5, 6}).
This is likely the pellet material reappearing in sector~16 after spreading around the torus.

From the recording for this discharge (\#41014), it appears that the material may come from both sides (co- and counter-current direction) as the radiating region does not appear to rotate in only one direction indicated by the arrows. It seems to rotate in co-current direction first, followed by a -- potentially dominating -- ctr-current rotation in~(f)--(h). Note, that potential aliasing or reflection effects are not taken into account for this description.

As the fragments enter the plasma, some of the outer flux surfaces might break-up and part of the thermal energy is lost conductively/convectively as observed in JOREK simulations described by W.~Tang~\etal~\cite{Tang_2025, Tang_2026}.
The strong radiation during the MFA will lead to an additional loss of plasma pressure. As the position control of the discharge control system~(DCS) can not react on this timescale (\mbox{$< 1$~ms}), this loss of pressure ($\beta$) results in a movement of the plasma towards the central post (see ``Rcurr'' signal~\& circle-marker in \subfig{PME}{b, f}) and together with a changing value of the plasma internal inductance~($li$), the plasma shape also changes (triangular-markers and corresponding signals in~(b, c)). At the same time, a strong radiation source is visible in the UHS recordings between the upper divertor and the central post, shown in \subfig{disruption_phases}{7--9} labeled as ``post injection''.

\subsection{Plasma Movement Event (PME) \label{sec:PME}}
\begin{figure*}
	\centering
	\includegraphics[width=\textwidth]{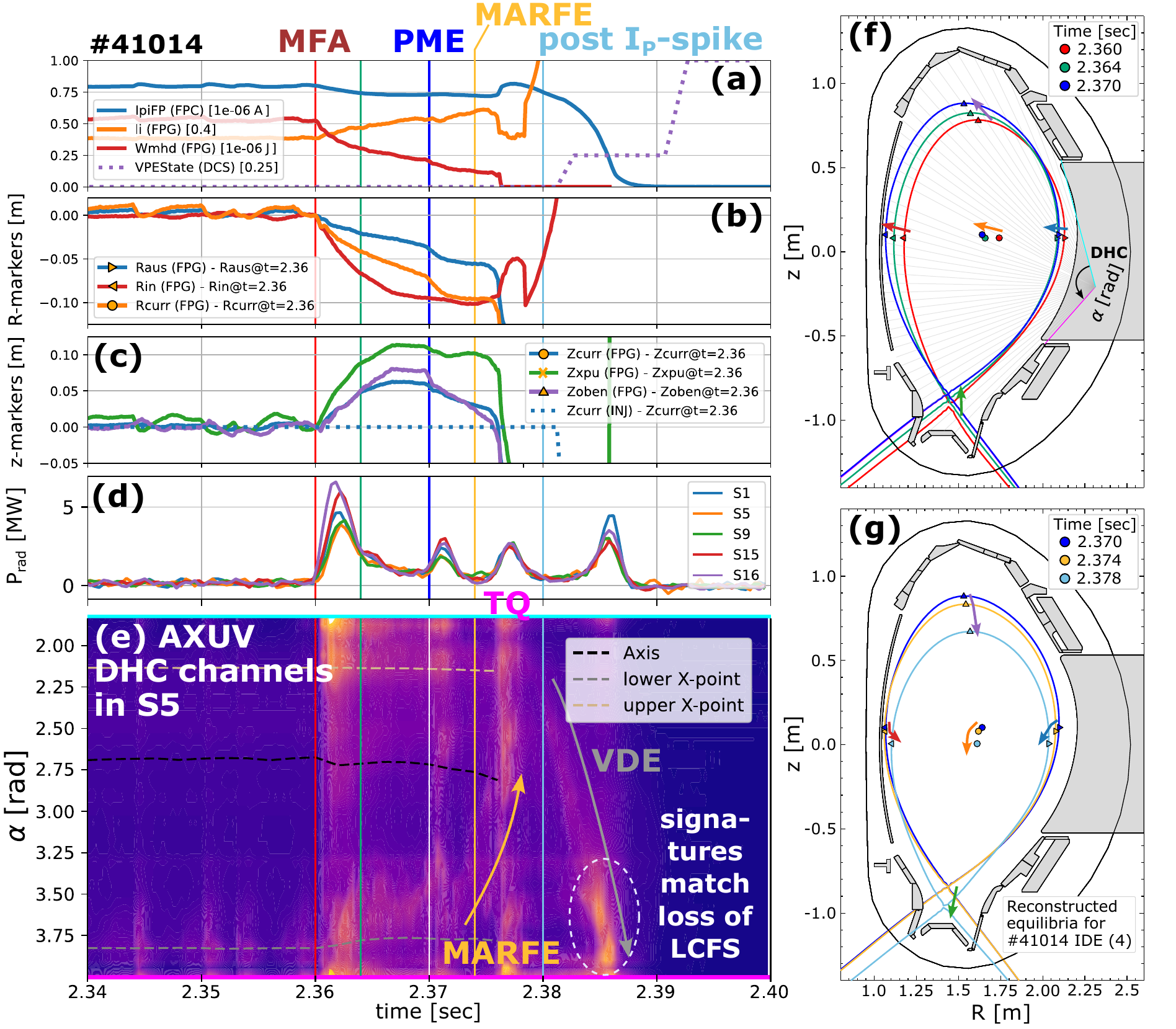}
	\caption{The plasma movement event in discharge \#41014: (a)~IP, plasma internal inductance~($li$), $W_\textrm{th}$, and \texttt{VPEState}, (b)~R-position markers, (c)~z-position markers, (d)~$P_\textrm{rad, sector}$. In~(e), a 2D plot of the \texttt{DHC} AXUV LOSs shown in~(f) is presented. The arrows in~(f, g) indicate the movement of the different control positions of the plasma, where the colour is the same as in (b,~c). (f)~Directly after the injection, the plasma moves closer to the HFS (central post) and the plasma shape changes, which can be connected to the initial pressure loss and changes in $li$. Note, that the DCS will also react on these changes of~$\beta$~\&~$li$, however, this does not involve a \texttt{VPEState}-change. At 2.370~s (10~ms after injection), the PME takes place: (g)~the movement of the $Z_\textrm{curr}$ is now directed downwards (towards the active X-point) and further towards the HFS while the plasma shape is also slightly changing. After the IP-spike, the plasma starts to move downwards (towards active X-point) during the VDE phase. The maximum of the last radiation peak (VDE peak) in~(d) appears around the time when the shunt currents drop (compare \subfig{disruption_phases}{a} around time frame~$\approx 22$) and the signatures match the loss of the last closed flux surface~(LCFS)~\cite{Artola_2024}. The reconstructions from \texttt{FPC/G} are calculated via function parametrisation~\cite{Braams_1986_FP, McCarthy_1992_PhD, Mc_Carthy_1999}.\label{fig:PME}}
\end{figure*}

For long pre-thermal quench~(pre-TQ) durations, caused by 100\%~$\textrm{D}_2$ or neon doped pellets, in some cases a second radiation peak is observed before the TQ (see \subfig{disruption_phases}{10--12}, \ref{fig:PME}(d), and \ref{fig:disruption_evolution_2}). As it is accompanied by the movement (and change in shape) of the plasma, we will refer to this event as the plasma movement event~(PME) and the radiation peak as the ``PME (radiation) peak'' throughout this paper.

During the PME, the current centroid~($Z_\textrm{curr}$) position is moving downwards  (towards the active X-point) and further towards the central post as the plasma shape also changes (\subfig{PME}{f}). This motion is also visible in the fast camera recordings (see \subfig{disruption_phases}{10--12}) as the radiation pattern moves further down the central post with increasing amplitude in the visible range compared to~\subfig{disruption_phases}{9}.
As the real-time $Z_\textrm{curr}$ reconstruction~(DDS, compared to FPG shot-file) does not exceed the 7~cm threshold in this discharge (not shown in \fig{PME}), the system state of the VPE in the DCS~(\texttt{VPEState}) remains unchanged, which indicates that this is not due to the DCS requesting e.g. a controlled ramp-down of the plasma~\cite{Heinrich_2025_PhD, Kudlacek_2024_DCS, Sieglin_2023_DCS}.
Only after the plasma current spike (IP-spike) had occurred, the \texttt{VPEState} changes in~\subfig{PME}{a} and a new current z-position (Zcurr -- INJ) is requested shown in~(c). Consequently, this movement of the plasma during the PME is not initiated by the DCS.

There is strong evidence, that a break-up of a few central flux surfaces may cause this event. 
During the pre-TQ phase, $li$ is steadily increasing (\subfig{PME}{a}). However, at the time of the PME, a small drop in $li$ is visible and the plasma current is increasing (better visible in \subfig{disruption_evolution_2}{e}). This is accompanied by a small drop in $W_\textrm{th}$. These signatures are the same that we observe on a bigger scale a bit later during the TQ: here, a full stochastisation of the field lines takes place and $W_\textrm{th}$ fully collapses.

This break-up of a few central flux surfaces during the PME will reduce the plasma pressure, hence lead to the further shift of the plasma towards the central post observed during this time~\cite{Gruber_1993, Lukash_2005}.
The thermal energy that escapes from the core plasma will then cause increased heating of the outer plasma leading to more efficient radiation of impurities in the scrape-off layer~(SOL), hence the second radiation peak.

\begin{figure*}
	\centering
	\includegraphics[width=\textwidth]{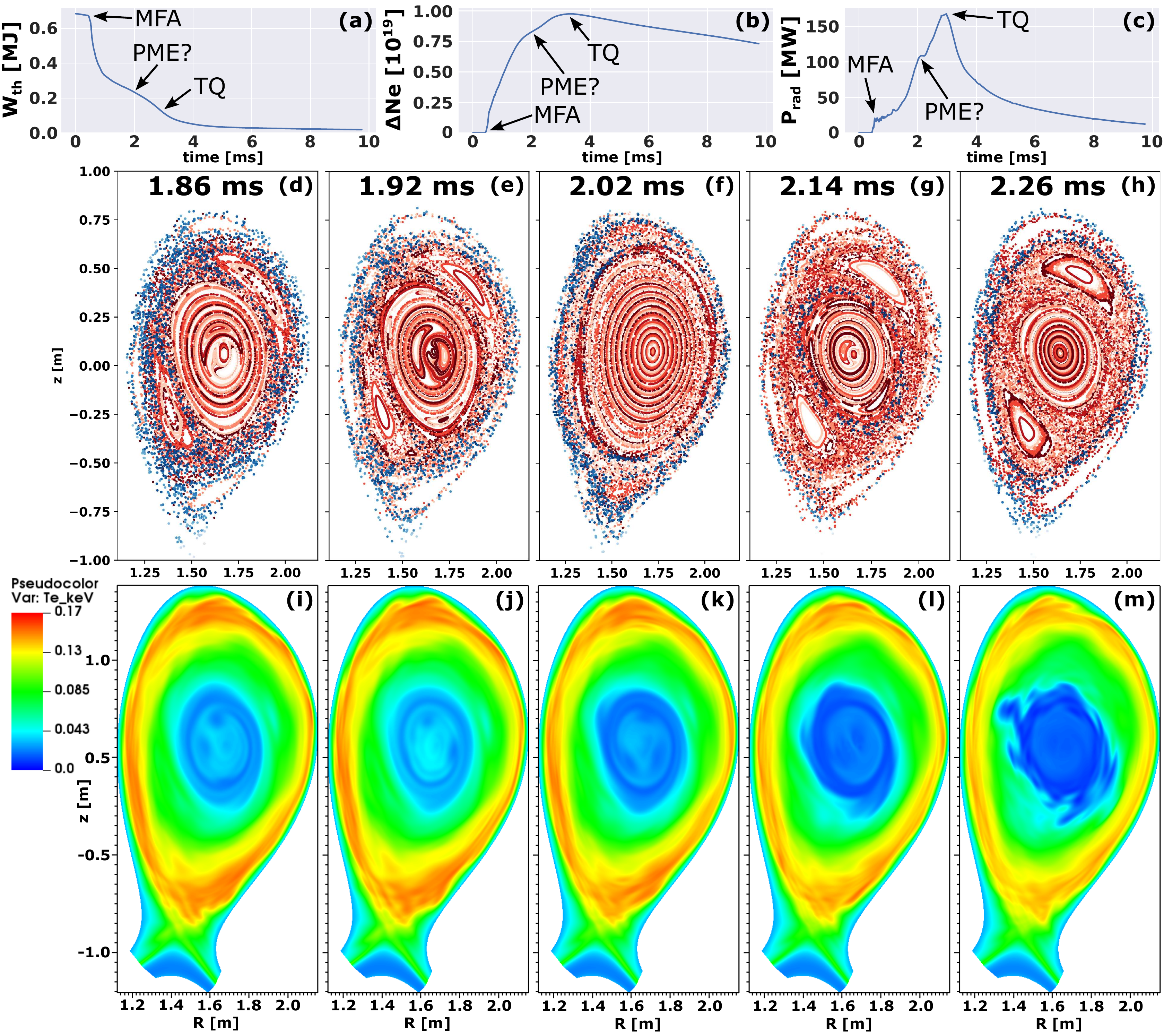}
	\caption{(1,~1)~kink mode that occurs in the JOREK simulations after a $\text{f}_\textrm{neon} = 0.12\%$~Ne injection (``SF FV Ne0.12'' from the paper by W.~Tang~\etal~\cite{Tang_2025}). The timing (between MFA and TQ) but also the signatures might indicate a connection to the experimentally observed PME.
	In~(a) $W_\textrm{th}$, (b)~neon assimilation $\Delta \textrm{Ne}$, and~(c)~total radiated power. The Poincar\'e plots \mbox{in~(d)$\rightarrow$(h)} indicate the flux surface break-up with the $T_\text{e}$ during the MHD activity \mbox{in~(i)$\rightarrow$(m)}.
	Whether the PMEs are caused by (1,~1) kink modes requires further studies for confirmation. \label{fig:JOREK_PME}}
\end{figure*}

The exact cause of this flux surface break-up is at present not yet fully understood, however, first indications hint towards the material assimilation processes that increase the plasma density strongly around that time which induce a strong magneto-hydrodynamic~(MHD) activity.
In the JOREK simulations for low neon contents shown in figure~3 of the paper by W.~Tang~\etal~\cite{Tang_2025}, a small radiation peak between fragment arrival and TQ is visible. In the case of the 0.12\%/1\% neon cases it occurs at about 2--3~ms after the injection, while the pellet material is still being assimilated in the plasma. At this time in the simulations, a (1,~1)~kink mode (toroidal and poloidal mode numbers \mbox{$n = m = 1$}) is present in the core region, which will lead to the partial stochastisation of central flux surfaces, shown in \fig{JOREK_PME} for the case ``SF FV Ne0.12'' from the paper by W.~Tang~\etal~\cite{Tang_2025}. This MHD event will decrease the electron temperature~(Te), hence the plasma pressure in the core region. At the same time, a small radiation peak is visible in~\subfig{JOREK_PME}{c} before the full stochastisation takes place during the TQ.
This would match the experimental observation of the PME.
Experimentally, coherent mode structures with $n=1$ and $n=2$ can be identified both in the magnetics and soft X-ray before the SPI injection and briefly during the quench, but no coherent mode structure can be identified between the two.

As indicated in figure~4 in the paper by A.~Patel~\etal~\cite{Patel_2025}, the time of the fragments reaching the core region is in the order of a few ms. The case shown in figure~4 is for a pellet with a velocity of 200~m/s, while $\text{v}_\textrm{pellet} = 466$~m/s ($25^{\circ}$, rectangular head) for \#41014, hence the fragments reach the $q = 2$ rational surface even earlier than the 2~ms. 
However, as shown in figure~19 in that paper, the density close to the magnetic axis does not start to increase before 8~ms after fragment arrival for a 100\%~$\textrm{D}_2$ injection with 270~m/s. This delayed increase is attributed to the comparably slow time scale of the diffusive transport of heat and particles after the material is deposited around $\Psi_\text{N} = 0.8$, 3~ms after the injection~\cite{Patel_2025}.
Therefore the strong increase in the central plasma density is a promising candidate for the trigger of this event.

The PME in our reference discharge \#41014, takes place rather late after the fragment arrival. Earlier times of the PME -- around 5~ms and earlier -- are also observed.

\subsection{MARFE \label{sec:MARFE}}
Directly after the PME for discharge \#41014, a MARFE~(multifaceted asymmetric radiation from the edge)~\cite{Lipschultz_1984} is observed to move upstream. 
The phenomenology of the MARFE~\cite{Lipschultz_1984, Lipschultz_1987_MARFE, Drake_1987, Stacey_1996, Stroth_2022_XPR_MARFE} can be understood as a sort of radiation condensation effect: a local increase in plasma density will lead to increased radiation in that area, which in return causes an increase in plasma density to maintain the local pressure balance~\cite{Drake_1987, Stacey_1996}. In diverted plasmas, this region will form close to the X-point (also referred to as ``X-point radiator~(XPR)''~\cite{Stroth_2022_XPR_MARFE, Sieglin_2024_MARFE, Sieglin_2024_SOFT}). With increasing density, the MARFE can start moving upwards along the high field side~(HFS)~\cite{Stroth_2022_XPR_MARFE, Lipschultz_1987_MARFE, Sieglin_2024_MARFE, Sieglin_2024_SOFT}. This movement can be stopped if enough heating is supplied, making it stationary in the new position~\cite{Sieglin_2024_MARFE, Sieglin_2024_SOFT}.
Once the radiating region has reached a far upstream position, it moves towards the LFS and deeper into the confined region (towards the magnetic axis) triggering the consequent disruption~\cite{Sieglin_2024_MARFE, Sieglin_2024_SOFT}. For AUG it was shown, that with the real-time actuators of the DCS, the displacement of the MARFE can be controlled and the disruption avoided~\cite{Kudlacek_2024_DCS}. Note, that in the experiments described in this paper, no such disruption avoidance action is taken. This way, we aimed to minimise the influence of the DCS on the disruption evolution.

The movement of the MARFE can be tracked with the help of the fast AXUV diodes as shown in \subfig{PME}{e} and is even visible in the fast camera recordings figure~\mbox{\ref{fig:disruption_phases}(14--16)}. It starts close to the X-point and moves \mbox{to--~and} upwards the central post. The sketch of the full MARFE trajectory is based on the tracking for \#37495~\cite{Klossek_2026_PhD}. It is only visible in a few LOSs at the same time, while the PME radiation peak shortly before was visible in all LOSs simultaneously.

\subsection{Thermal quench (TQ) and IP-spike \label{sec:TQ}}
If the injected material is sufficient to cause a thermal quench, the TQ phase is initiated. For 100\%~$\textrm{D}_2$ injections, the disruption threshold is located around $2\times 10^{22}~\textrm{D}_2$~atoms~\cite{Heinrich_2025_frad, Hoelzl_2020_D2-SPI}. Below this number, the SPI discharges were recovered by the DCS of AUG without the occurrence of a disruption, discussed in more detail in the paper by P.~Heinrich~\etal~\cite{Heinrich_2025_frad}.
The flattening of the plasma current profile during the TQ causes the characteristic plasma current spike (IP-spike) as the conservation of helicity requires an increase in the total plasma current to counteract these changes~\cite{Biskamp_1997, Nardon_2023}.
This IP-spike can be up to 10\% of the initial plasma current~\cite{Mirnov_1998, Hender_2007_ITER_Physics, Nardon_2023} and depends on the change of the plasma current profile~\cite{Hender_2007_ITER_Physics, Nardon_2023, Boozer_2019c, Boozer_2020a}. Consequently, the quantity and kind of material added to the plasma (e.g. impurities for disruption mitigation) affects the changes of plasma resistivity~($\eta$) and with it the properties of the IP-spike (i.e. the height and width)~\cite{Hoelzl_2021, Nardon_2021a, Nardon_2023}.

\begin{figure*}
	\centering
	\includegraphics[width=\textwidth]{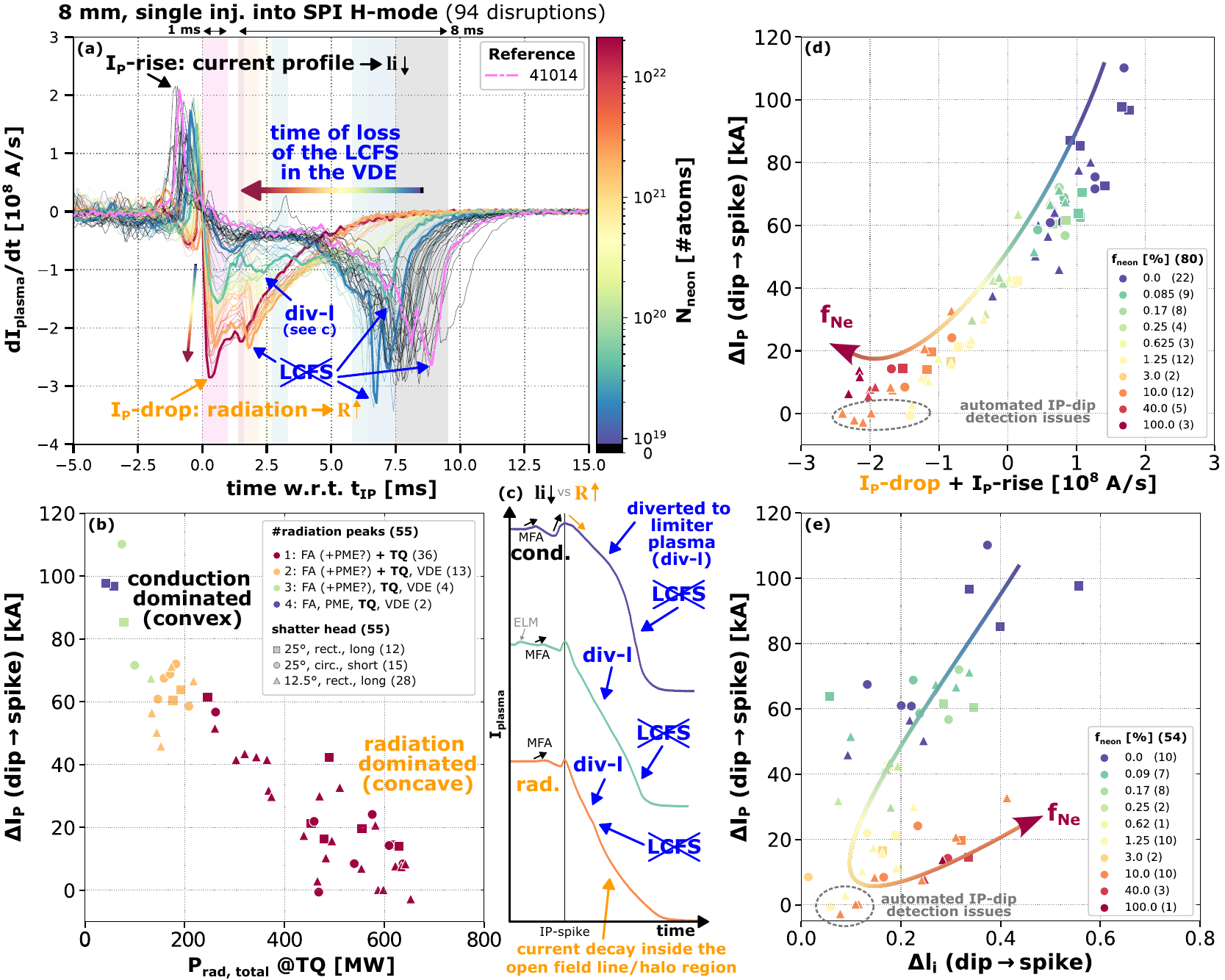}
	\caption{IP-spike height changes due to competing $li$ and plasma resistance~(R). In~(a), the time evolution of the $\textrm{dI}_\text{plasma}/\textrm{dt}$ is given. In~(b), The IP-spike height as a function of the total radiated power at the radiation peak closest to the TQ. The color coding indicates how many radiation peaks were detected. Note, for less than 3 detected peaks, the radiation from the MFA + TQ is used as the TQ alone is not resolved. In~(c), the transition from diverted to limiter plasma~(div-l) and the loss of the LCFS (see section~\ref{sec:Evolution}) during the VDE is shown -- visible as the $\textrm{dI}_\text{plasma}/\textrm{dt}$ minimum during the VDE in~(a) and maximum of the shunt currents ($\rightarrow \textrm{I}_\text{halo}$) in~\ref{fig:disruption_evolution_2}(e). In~(d) the height of the IP-spike is plotted as a function of the $\textrm{I}_\text{P}$-rise (maximum pre-IP-spike) and $\textrm{I}_\text{P}$-drop (local minimum post-IP-spike) peaks of $\textrm{dI}_\text{plasma}/\textrm{dt}$. In~(e), the height is plotted as a function of the $li$-drop during the TQ. For similar values of $\Delta li$, different IP-spike heights are observed, depending also on the post-TQ plasma resistance. \label{fig:IPspike_dIp_dt}}
\end{figure*}
For long disruption times, as shown in \subfig{disruption_phases}{b}, a third radiation peak (``TQ (radiation) peak'') is visible during this phase in the foil bolometers, which is usually toroidally symmetric in these cases. The cause of this radiation peak is most likely similar to the one suspected for the PME described above, where the stochastisation of the field lines leads to an enhanced heat transport into the SOL, thus increasing radiation. The IP-spike height and width also depends on the neon content of the injections, as this strongly affects the TQ dynamics and post-TQ resistivity. Hereby, the higher the assimilated neon content, the smaller the amplitude and the width of the IP-spike as shown in figures~\ref{fig:disruption_evolution_1} and~\ref{fig:IPspike}. This connection of the IP-spike height to the neon content may be explained by the transition from conduction to radiation dominated disruptions. With increasing radiated power around the time of the TQ -- either by increasing the estimated number of neon atoms~($\text{N}_\textrm{neon}$) or changing shattering parameters -- an additional local minimum besides the VDE-minimum develops in $\textrm{dI}_\text{plasma}/\textrm{dt}$ after the IP-spike, labeled as ``$\textrm{I}_\text{P}$-drop'' in~\ref{fig:IPspike_dIp_dt}(a). The increased radiated power increases the resistance of the plasma, consequently causing a stronger plasma current decay. Therefore, the  IP-spike height results from the battle between 
\begin{itemize}
	\item the ``$\textrm{I}_\text{P}$-rise'': changing current profile $\rightarrow li{\downarrow}$ and $\Delta li{\uparrow}$ in figure~\ref{fig:IPspike_dIp_dt}(e) and 
	\item the ``$\textrm{I}_\text{P}$-drop'': increasing radiation $\rightarrow \text{R}{\uparrow}$ in figure~\ref{fig:IPspike_dIp_dt}(b). 
\end{itemize}	
For high neon content in the pellet (radiation dominated disruptions), the $\textrm{I}_\text{P}$-drop becomes larger as the $\textrm{I}_\text{P}$-rise for conduction dominated/unmitigated disruptions, hence the IP-spike height decreases indicated by figures~\ref{fig:IPspike_dIp_dt}(d) and~\ref{fig:IPspike}. As indicated by figure~\ref{fig:IPspike_dIp_dt}(e), the $\Delta li$ alone seems to be insufficient to explain the IP-spike height as the same $\Delta li$-values can be seen for different values of $\Delta I_\textrm{p}$. Additionally, the temporal evolution of $li$ during the VDE phase changes between conduction and radiation dominated disruptions. Poorly mitigated disruptions, behave similar to hot VDEs, where $li$ increases at the transition to a limiter plasma (see figure~\ref{fig:PME}(a) at \mbox{$t=2.38$s}), as more and more current is peeled off. This also causes the ``unphysical trend'' in figure~\ref{fig:scenario}(I) with \mbox{$W_\textrm{mag} = L\cdot I_\textrm{p}^2 = \mu_0 \cdot \text{R} \cdot \left[ln(8\text{R}/\text{a}) - 2\right] \cdot I_\textrm{p}^2$~\cite{Zohm_1993}}, with major radius~R and minor radius~a. However, for radiation dominated disruptions, $li$ strongly decreases soon after the IP-spike, when the LCFS is lost and the entire current now flows through the open field line region.

\subsection{Current quench (CQ) and Vertical displacement event (VDE) \label{sec:CQ_VDE}}
The early current quench~(CQ) rates with unscaled duration of the CQ~($\Delta \textrm{t}_\textrm{CQ}^{100 \rightarrow 80}$) were used as an approximation for the assimilated neon in our SPI studies~\cite{Jachmich_2023_EPS, Jachmich_2024_EPS}: higher amounts of assimilated neon --~hence radiation dominated disruptions~-- will cause faster early CQ rates, leading to the characteristic exponential decay curve of the plasma current (concave-from-below).
For 4~mm pellets, we frequently observed multiple ``shoulders'' during the CQ (see appendix~A.11 in the PhD thesis by P.~Heinrich~\cite{Heinrich_2025_PhD}), which may be connected to differences in the VDE phase.

In case of injections with $\text{f}_\textrm{neon} < 0.17$\% (CQ-shape~(1) in \fig{disruption_evolution_2}), the early CQ rate is low. The motion of the plasma during the VDE is accelerated throughout the CQ~\cite{Schwarz_2024_PhD, Gruber_1993, Boozer_2019}. After the transition from a diverted plasma to a limiter plasma~(``div-l'' in figures~\ref{fig:disruption_phases} and~\ref{fig:IPspike_dIp_dt}) as the X-point touches the wall, more and more layers of the plasma are scraped-off and \mbox{$\lvert \textrm{d}I_\textrm{p}/\textrm{dt}\rvert$} will increase, creating the overall convex-from-below CQ shape as illustrated in figures~\ref{fig:PME} and~\ref{fig:disruption_evolution_2}(e).
This vertical movement towards the lower divertor (in direction of the active X-point) is visible in the fast camera recordings shown in \subfig{disruption_phases}{19--22}. While for the 8~mm, 100\%~Ne SPI (\#40679) an almost smooth current decay is visible, a bigger ``shoulder'' in the plasma current or change in the decay rate (see figures~\ref{fig:IPspike_dIp_dt}(a, c) and~\ref{fig:disruption_evolution_2}(c)) is observed for intermediate to high neon content injections. This phenomenon is linked to the phase of the VDE where the LCFS is lost as the magnetic axis touches the wall~\cite{Artola_2024}, and a drop in halo currents is observed in the experiments and JOREK VDE simulations~\cite{Artola_2024, Schwarz_2023, Schwarz_2024_PhD}. The remaining current now decays in the open field line/halo region and while the magnetic axis left the domain, the current axis moves from the midplane upwards again~\cite{Schwarz_2023, Schwarz_2024_PhD}.
Note, that in these investigations, VDEs are leading to the disruptions after intentionally displacing the plasma towards the divertor in the experiments~\cite{Schwarz_2023, Schwarz_2024_PhD}, instead of ``natural'' VDEs that typically occur after the TQ in the SPI experiments discussed in this paper.
Around the loss-time of the last closed flux surface~($\textrm{t}_\textrm{LCFS}$), the maximum of the last radiation peak (``VDE peak'') is observed for injections with $\text{f}_\textrm{neon} < 0.17$\% (CQ-\mbox{shapes~(1--2)}), shown in figures~\ref{fig:disruption_phases}(22), \ref{fig:PME}(e) and~\ref{fig:disruption_evolution_2}(e). For injections with $\text{f}_\textrm{neon} > 0.17$\% (CQ-shapes~(3--4)), a small radiation plateau is observed until loss-time of the last closed flux surface~($\textrm{t}_\textrm{LCFS}$), shown in \ref{fig:disruption_evolution_2}(b). This evolution of the CQ-shape is described in more detail in the following.

\section{Disruption Evolution \label{sec:Evolution}}
\begin{figure*}
	\centering
	\includegraphics[width=\textwidth]{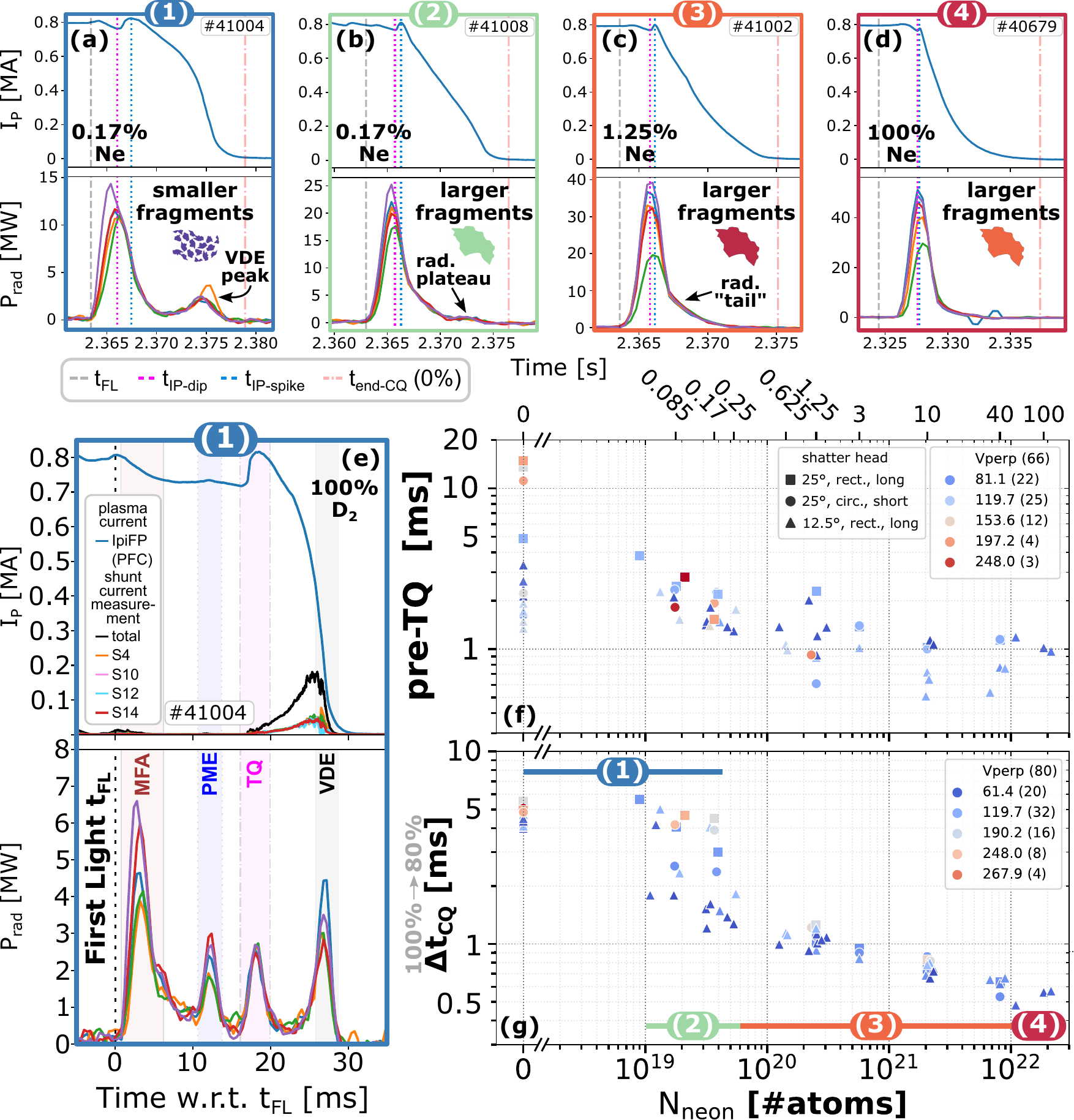}
	\caption{Evolution of the disruption dynamics with increasing amount of assimilated neon content. From~(1) to~(4), the amount of assimilated neon increases and a transition from convex to linear and finally concave CQ shapes is visible. At the same time, the VDE peak is reducing: first the amplitude decreases until it looks more like a radiation plateau shown in~(b). Afterwards, this plateau looks more like a smooth ``tail'' of the single radiation peak.
	The pellets launched in~(a) and~(b) share the same neon content but with varied $\text{v}_\perp$, hence fragment size distributions, illustrating the importance of the fragment size distributions in the transition from convex to linear CQ shape. This is also visible as lower values of $\text{v}_\perp$
	In~(f) the pre-TQ duration and in~(g) the early CQ duration ($\Delta \textrm{t}_\textrm{CQ}^{100 \rightarrow 80}$) as a function of $\text{N}_\textrm{neon}$ are shown. \label{fig:disruption_evolution_2}}
\end{figure*}

As demonstrated in the previous section, the properties of the disruptions and even the event chain depends strongly on the amount of assimilated neon in the plasma.
The figures~\ref{fig:disruption_evolution_1} and~\ref{fig:disruption_evolution_2} show the disruption evolution as a function of injected neon content.
While the injection parameters like fragment size and velocity distributions affect the efficiency of the neon assimilation, the total amount of injected neon is the dominant parameter here.
The effect of the injection parameters, which are especially relevant for the disruption evolution in the low neon doping range, are discussed in the next section.

In the following, the disruption evolution is described for increasing neon content in the pellet, starting at 100\%~$\textrm{D}_2$ injections and ending at 100\%~Ne injections following the CQ-shapes from \fig{disruption_evolution_2}:

\begin{itemize}
\item[{\parbox[t]{0.1\linewidth}{(no dis-rup-tion)}}] \parbox[t]{1\linewidth}{For 100\%~$\textrm{D}_2$ injections below $2\times 10^{22}$ deuterium atoms, the DCS-settings of AUG for the 2022 experiments can recover the discharge and bring it back to pre-injection parameters~\cite{Heinrich_2025_frad}. Only a small, toroidally symmetric radiation peak is observed as the main fragments arrive.}

\item[(1)] In case of an injection of a full-length, 8~mm, 100\%~$\textrm{D}_2$ pellet or neon fractions up to 0.085\% (8~mm) / 1.25\% (4~mm), the plasma will disrupt and up to four radiation peaks can be observed in case of a long disruption time scale (see \subfig{disruption_evolution_2}{e}). These radiation peaks correspond to the time windows of the MFA, PME, TQ, and VDE, respectively. For these disruptions, the CQ-shape is convex and IP-spike-amplitudes of up to 105\% of the average pre-injection plasma currents ($\approx 115$\% with respect to minimum in the plasma current before the IP-spike~(IP-dip) are observed, see figure~\ref{fig:IPspike}.

\item[(1--2)] For increasing amount of assimilated neon, hence shorter pre-TQ durations, the observed radiation peaks merge --~due to the time resolution of the foil bolometers~-- and up to two radiation peaks can be observed:\\
MFA (+ PME?) + TQ as well as a separated peak during the VDE phase (compare \#41004 in \subfig{disruption_evolution_2}{a}). The presence of the PME phase for these disruptions is not confirmed.

\item[(2)] The higher the assimilated amount of neon (compare \subfig{disruption_evolution_2}{b}), the more linear the current quench becomes. At the same time, the pre-TQ duration and the amplitude of the IP-spike are further reduced.
Additionally, the second radiation peak (VDE peak) is reduced with increasing neon assimilation until it forms a radiation plateau -- slightly above the pre-injection radiation level -- which lasts until the loss of the $\textrm{t}_\textrm{LCFS}$ or ``hot core'' (explained later in this section), when the plasma current also approaches zero.

\item[(3--4)] For even higher neon fractions/assimilation (\subfig{disruption_evolution_2}{c}), the CQ starts to exhibit a concave shape with a prominent shoulder as introduced in \sect{CQ_VDE}. The radiation plateau vanishes, thus the radiation tail directly approaches zero after the single radiation peak.
The prominent CQ shoulder almost vanishes for the 100\%~Ne SPI.
\end{itemize}

From 100\%~$\textrm{D}_2$ to 100\%~Ne injections, the relative asymmetry of the radiation between \mbox{S16} and the neighbouring sectors reduces, while the asymmetry of sectors~5 and~9 increases~\cite{Heinrich_2025_PhD}.

In summary, the CQ shape transitions from a convex to a linear and finally a concave shape, while the four distinct radiation peaks turn into a single one and the halo currents (indicated by shunt currents in \subfig{disruption_evolution_1}{a) $\rightarrow$ (f}) are reduced.
A similar observation of linear~(2)/concave~(3--4) CQ-shapes had been related to the presence/absence of VDEs in figure~2 by R.~Yoshino~\etalc{Yoshino_1996}, respectively. Hereby, the fast shut-down of the plasma via injection of a neon ``killer-pellet'' (compare~(3--4)) stabilised the plasma position~\cite{Yoshino_1996}.
This transition of the CQ-shape may be explained the following way:

In radiation dominated disruptions, the plasma volume remains approximately constant owing to the presence of broad halo currents, which stabilise the current centroid~($\textrm{Z}_\textrm{curr}$) near the midplane even if the plasma core~($\textrm{Z}_\textrm{mag}$) is displaced~\cite{Schwarz_2023, Schwarz_2024_PhD}. At the same time, $T_\text{e}$ becomes comparable in the core and edge regions (few to 10s of eV)~\cite{Artola_2024}, resulting in a relatively flat resistivity profile. Under these conditions, the effective plasma resistance can be treated as nearly constant, so that the plasma current decays exponentially, as expected from the standard \textit{RL}-circuit description. This explains the characteristic concave shape of the current quench.

On the other hand, for conduction dominated disruptions, the core plasma remains hotter than the SOL. Consequently, the resistivity profile moves along with the plasma~$\textrm{Z}_\textrm{mag}$. With the core resistance being $\propto \textrm{a}^{-2}$ (for constant resistivity), the resistance increases and the CQ accelerates ($\rightarrow$~convex shape) when the plasma size is reduced during the VDE.

The intermediate situation is what may explain the linear CQ-shapes: the plasma is not fully collapsing and a smaller hot region remains around the magnetic axis. When the plasma starts to move, part of the current can be reinduced in the hot region~\cite{Artola_2024} (effectively slowing down the decay) until the flux surface encapsulating the ``hot core'' hits the plasma facing components~(PFCs) and the CQ is again accelerated (see figure~\ref{fig:disruption_evolution_2}(b) at~$\sim 2.373$~s), this time causing no visible peak in radiation. Here, higher neon contents in the pellet~($\text{f}_\textrm{neon}$) and larger fragments (penetrating deeper into the plasma) lead to more linear CQes until no hot plasma core remains, the resistivity profile becomes flat and the concave CQ-shape is observed.

Note, that already for a small amounts of neon in the pellets, the TQ is triggered too quickly after the fragment arrival to resolve the first three radiation peaks as separate peaks in the foil bolometers. Therefore, the presence of the PME and the contribution of the MFA, PME, and TQ to the total radiation is currently undetermined.

The reduction of the shunt currents and with it the VDE radiation peak -- with the maximum at $\textrm{t}_\textrm{LCFS}$ -- in our experiment are in line with previous observations: 
increasing the amount of injected/assimilated neon, the mitigation efficiency of the otherwise violent VDEs is expected to increase, causing lower shunt currents in the process~\cite{Yoshino_1996, Artola_2024, Schwarz_2023, Schwarz_2024_PhD}.
Additionally, the amplitude of the IP-spike is also reducing with increasing neon content (see figures~\ref{fig:IPspike_dIp_dt} and~\ref{fig:IPspike}). The CQ shape, as it is directly linked to the neon assimilation, can already be used as a first indicator of the mitigation efficiency on the shot day from the plasma current measurement alone and especially relevant in the absence of valid density measurements.

\begin{table*}[b]
        \centering
            \begin{tabular}{c|c|c|c|c|c}
            \toprule\specialrule{0.5pt}{1.5pt}{\belowrulesep}
                \multicolumn{6}{c}{pre-TQ and CQ durations} \\
                \hline
                 & $\Delta \textrm{t}_\textrm{pre-TQ}$ & \dtCQ{100}{0} & \dtCQ{100}{80} & \dtCQ{80}{20} & \dtCQ{90}{10}\\
                \hline
                injected material & [ms] & [ms] & [ms] & [ms] & [ms]\\
                \hline
                \hline
                \phantom{0.}100\% $\textrm{D}_2$ & 1.3 -- 14.9 & 10.7 -- 13.3 & 4.0 -- 5.5 & 3.0 -- 5.4 & 5.2 -- 9.0 \\
                \hline
                0.085\% Ne & 1.5 -- \phantom{1}3.8 & \phantom{1}9.7 -- 12.2 & 1.8 -- 5.6 & 2.8 -- 5.1 & 4.9 -- 6.6\\
                \hline
                \phantom{0.0}10\% Ne & 0.5 -- \phantom{1}1.1 & \phantom{1}8.2 -- 11.1 & 0.7 -- 0.9 & 2.7 -- 3.7 & 3.9 -- 5.1\\
                \hline
                \phantom{0.}100\% Ne & 1.0 -- \phantom{1}1.2 & \phantom{1}9.6 -- 10.2 & 0.5 -- 0.6 & 2.7 -- 2.8 & 4.2 -- 4.3\\
            
            \specialrule{0.5pt}{\aboverulesep}{1.5pt}\bottomrule
            \end{tabular}
            \caption{Duration of the pre-TQ and CQ for injections with different $\text{f}_\textrm{neon}$ taken from \subfig{disruption_evolution_2}{f, g}. All CQ durations are unscaled (e.g. \dtCQ{80}{20} is not multiplied by 5/3). With increasing $\text{f}_\textrm{neon}$, the pre-TQ and CQ durations decrease. Already a minor neon doping of $\text{f}_\textrm{neon} = 0.085\%$ leads to a drastic reduction in the maximum pre-TQ duration to values below 4~ms (compare \subfig{disruption_evolution_2}{f}). \label{tab:CQ_durations}} 
\end{table*}

While the neon content in the pellet is the dominant parameter determining the maximum assimilation of neon in the plasma -- hence the disruption evolution -- the injection parameters (fragment size \& velocity) were shown to also affect the assimilation of neon~\cite{Jachmich_2023_EPS, Jachmich_2024_EPS}.
The most noticeable effects are present for low neon doping, in the overlap of~(1) and~(2), illustrated in figure~\ref{fig:disruption_evolution_2}(g). For the same neon content, better assimilation (also reflected in faster $\Delta \textrm{t}_\textrm{CQ}^{100 \rightarrow 80}$-rates~(g)) were reported before~\cite{Jachmich_2023_EPS, Jachmich_2024_EPS} for larger \& slower fragments -- produced by the $12.5^{\circ}$ head. As presented in this paper, these injections also show more linear CQes (compare~(a) with~(b)).

\section{Disruption timescales \label{sec:Time_scales}}

In this section, an overview of the time scales for the pre-TQ and CQ of the SPI induced disruptions at AUG is provided.

\begin{figure}[htb]
	\centering
	\includegraphics[width=\linewidth]{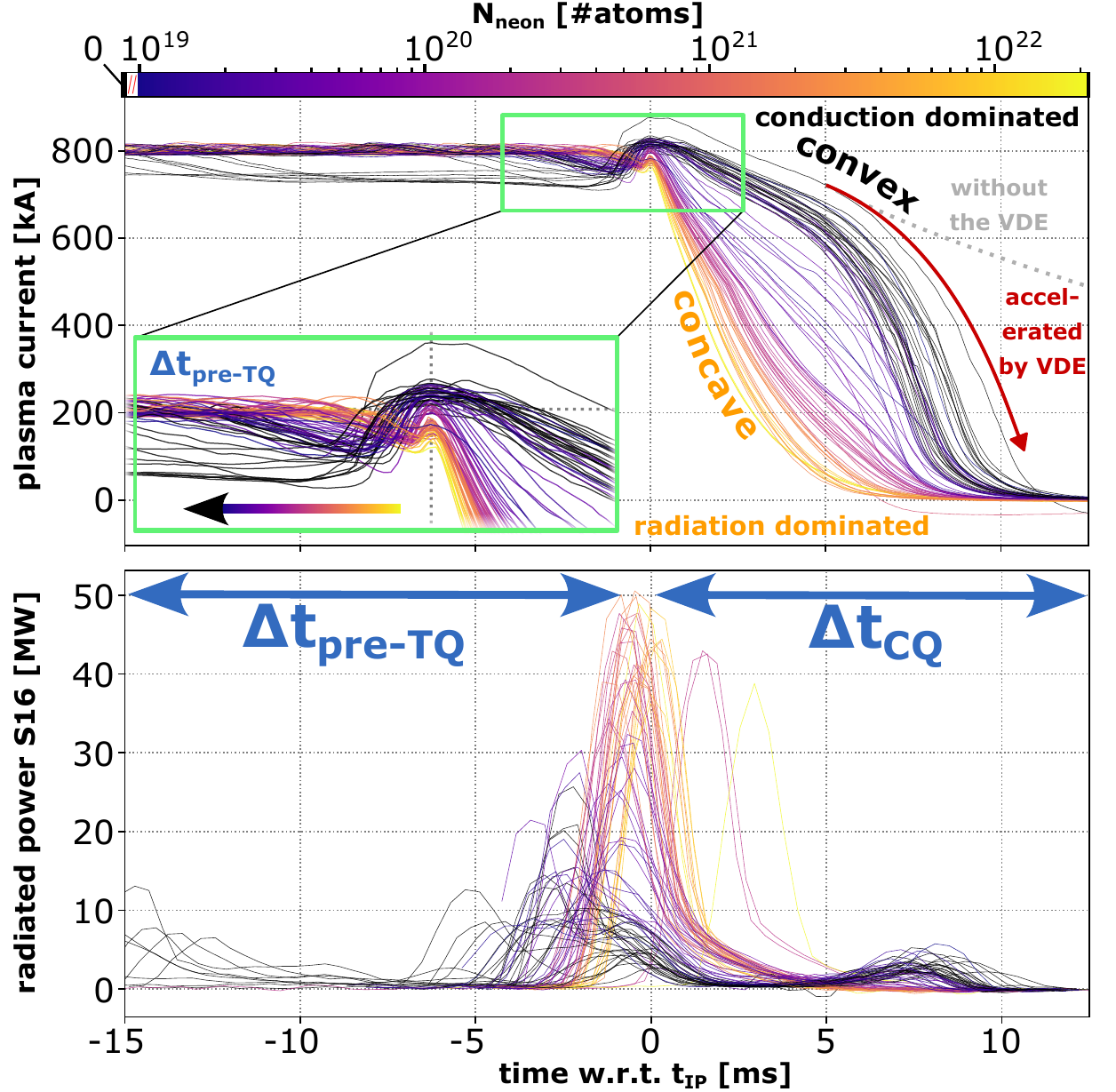}
	\caption{Continuous reduction of pre-TQ and CQ durations with increasing neon content, hence peak radiation. The (a)~plasma current and (b)~radiated power in \mbox{S16} are shown aligned with respect to the IP-spike. The transition from convex to linear and finally concave CQ shapes is observed in this continuous disruption evolution. \label{fig:disruption_lines}}
\end{figure}

With increasing neon assimilation the pre-TQ and CQ durations are decreasing, shown in \subfig{disruption_evolution_2}{f, g}, \subfig{disruption_lines}{a}, and \tabl{CQ_durations}.
This reduction is a continuous process accompanied by an increase in radiation, illustrated in \fig{disruption_lines}.

The $\Delta \textrm{t}_\textrm{pre-TQ}$-values are calculated from $\textrm{t}_\textrm{MFA}$ until the time of the IP-dip~($\textrm{t}_\textrm{IP-dip}$) (see \subfig{disruption_lines}{a}). Here, the time marker $\textrm{t}_\textrm{MFA}$ is used -- compared to the $\textrm{t}_\textrm{FL}$ -- as the exact timing of the MFA is easier to identify, hence more precise.

It had been shown for 100\%~$\textrm{D}_2$ SPI~\cite{Jachmich_2023_EPS, Jachmich_2024_EPS}, that $\Delta \textrm{t}_\textrm{pre-TQ}$ varies strongly with the injection parameters, around 1--15~ms. Hereby, disruptions with long pre-TQ durations \mbox{($> 10$~ms)} show four radiation peaks and disruptions with shorter $\Delta \textrm{t}_\textrm{pre-TQ}$ only show two or three radiation peaks.
Doping the pellets with 0.085\%~Ne already reduces the maximum pre-TQ duration significantly (\mbox{$< 4$~ms}), as was also observed in e.g. JET~\cite{Sheikh_2024_APS}. This may be a limiting factor for the planned staggered injection scheme for ITER as it brings two problems with it:
firstly, the pre-TQ durations might be too short to lead to high deuterium assimilation fractions, necessary to prevent the formation of high current RE-beams~\cite{Sheikh_2024_APS}.
Further, these reduced pre-TQ durations may require a high precision in the timing of the injections.

\begin{figure}[htb]
	\centering
	\includegraphics[width=0.5\textwidth]{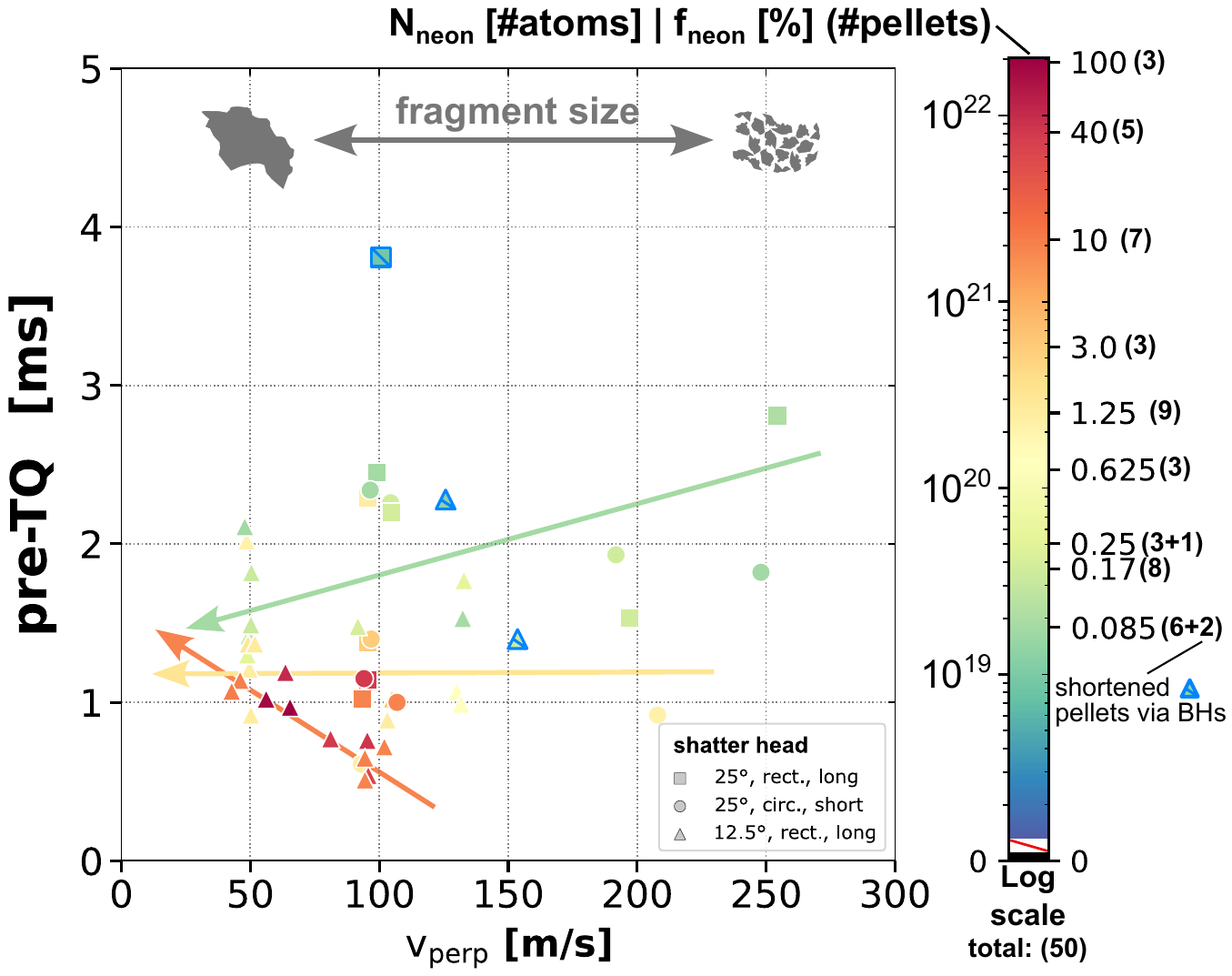}
	\caption{The pre-TQ duration as a function of the perpendicular velocity component for pellets containing neon. The perpendicular velocity component can be used as a approximation for the fragment size, however, the mean fragment velocity also changes with $\text{v}_\perp$, hence both effects have to be taken into account here. \label{fig:dt_vperp}}
\end{figure}

The values for the $\Delta \textrm{t}_\textrm{CQ}^{100 \rightarrow 80}$ shown in \subfig{disruption_evolution_2}{g} indicate the early CQ durations. This metric is beneficial for a cleaner inter-shot comparison -- compared to the often used metrics \dtCQ{80}{20} and \dtCQ{90}{10} -- as the early CQ rate is less influenced by VDE dynamics~\cite{Jachmich_2023_EPS, Jachmich_2024_EPS}.

In \fig{dt_vperp}, the pre-TQ duration is plotted as a function of $\text{v}_\perp$ -- used as an approximation of the mean fragment size. 
Important to note, however, for values of $\text{v}_\perp \gtrsim 150$~m/s, the mean fragment size does not change significantly anymore, whereas the mean $\text{v}_\textrm{fragment}$ still increases $\propto \text{v}_\textrm{pellet}$~\cite{Peherstorfer_2022_fragmentation, Heinrich_2025_PhD}.
In this empirical scaling, the mean fragment size~($\overline{d}_{\textrm{frag}}$)
\begin{align}
\overline{d}_{\textrm{frag}} &\propto \exp(-\text{v}_\perp)~\text{\cite{Peherstorfer_2022_fragmentation}}\\
\text{v}_\perp &= \text{v}_\textrm{pellet} \cdot \sin(\alpha))\\
\text{v}_\textrm{fragment} &\approx (\text{v}_\textrm{pellet} + \text{v}_\|) / 2 ~\text{\cite{Peherstorfer_2022_fragmentation}}\\
&= (\text{v}_\textrm{pellet} + \text{v}_\textrm{pellet} \cdot \cos(\alpha)) / 2\\
&= \text{v}_\textrm{pellet} \cdot (1+\cos(\alpha)) / 2
\end{align}
is not linearly proportional to $\text{v}_\textrm{pellet}$ unlike $\text{v}_\textrm{fragment}$. Hence, with increasing $\text{v}_\textrm{pellet}$, the ratio of $\text{v}_\textrm{fragment}/\overline{d}_{\textrm{frag}}$ increases exponentially.
In addition, larger values of $\text{v}_\perp$ are expected to also decrease the solid-to-gas fraction generated during the shattering process. This reduces the effective solid mass injected as the form of pellets and the gas may also affect the penetration characteristics of the material.  
These two effects may explain the potential roll-over effects, observed for the 100\%~$\textrm{D}_2$ injections reported before~\cite{Jachmich_2023_EPS, Jachmich_2024_EPS}. This is further supported by the observation of shorter pre-TQ durations for injections with the $12.5^{\circ}$, rectangular shatter head, producing larger~\& faster fragments at higher $\text{v}_\textrm{fragment}$~\cite{Jachmich_2023_EPS, Jachmich_2024_EPS}.
Additionally to this, we observed an apparent reversal of trends for the pre-TQ durations shown in~\fig{dt_vperp}: Below around 1.25\% neon, lower values of $\text{v}_\perp$ result in shorter pre-TQ durations, and above this value of $\text{f}_\textrm{neon}$ this trend reverses.
We observed a similar trend for $\text{f}_\text{rad}$~\cite{Heinrich_2025_frad, Heinrich_2025_PhD}: for neon concentrations above $\text{f}_\textrm{neon}$~$\sim 1.25\%$, higher values of $\text{v}_\perp$ seem to increase $\text{f}_\text{rad}$ in contrast to the neon doped cases. This would then match the observation of earlier thermal quenches, hence reduced pre-TQ durations.

With the data analysed so far, we propose the following explanation.
Larger fragments (lower values of $\text{v}_\perp$) can penetrate deeper into the plasma core -- potentially also linked to the plasmoid drift \& rocket force effects -- hence lead to increased assimilation of the material. In the case of pellets that contain neon, this leads to increased amounts of assimilated neon, hence shorter early CQ durations.
For the neon doped pellets with $\text{f}_\textrm{neon} < 1$\%, we believe that the TQ is triggered by radiation at the q~=~2 rational surface (helical cooling), hence the pre-TQ becomes shorter for larger \& faster fragments that would experience less plasmoid drift \& rocket force effects.
On the other hand, for pellets with high amounts of neon, larger \& faster fragments reduce the neon deposited at the plasma edge, hence prevent an early TQ due to a radiational collapse of this edge region. This would therefore delay the TQ and provide more time for the fragment material to assimilate in the core region, matching increasing durations of the pre-TQ while still matching the previously reported decreasing early CQ durations ($\Delta \textrm{t}_\textrm{CQ}^{100 \rightarrow 80}$)~\cite{Jachmich_2023_EPS, Jachmich_2024_EPS}.
Furthermore, this is supported by the observation of higher $\text{f}_\text{rad}$-values for higher values of $\text{v}_\perp$ (small \& fast fragments)~\cite{Heinrich_2025_frad, Heinrich_2025_PhD}. Here, the increased surface-to-volume ratio may cause higher peak radiation and the radiative collapse of the edge may explain the lower pre-TQ durations. At high neon fractions, the plasmoid drift is expected to be strongly suppressed, which may further support this transition in trends around $\text{f}_\textrm{neon} \sim 1.25\%$.

\section{Summary\label{sec:summary}}
The SPI-induced disruptions in the 2022 experimental campaign in ASDEX Upgrade follow a specific chain of events -- the so-called disruption phases: 

\begin{itemize}
\item The arrival of first, tiny fragments and gas (generated during the pellet break-up) visible in the fast AXUV signals is referred to as the first light~(FL).

\item This is followed by the main fragment arrival~(MFA), causing first large radiation peak detected by the foil bolometers. In this paper, it also marks the onset of the pre-thermal quench~(pre-TQ) phase.

\item For long pre-TQ durations an event referred to as the plasma movement event~(PME) is observed. Given the experimental signatures and matching observations in JOREK simulations, we believe this is caused by an MHD event that causes the break-up of a few central flux surfaces, thereby, reducing the plasma pressure and causing a shift of the plasma position and increased radiation ($2^\textrm{nd}$ radiation peak).

\item After this event, the presence of a MARFE can be observed for non-radiation dominated disruptions (e.g. 100\% $\textrm{D}_2$) and 
\item the TQ, followed by the characteristic IP-spike and CQ.
Hereby, the IP-spike height/width and early CQ rate, strongly depends on the amount of assimilated neon in the plasma and consequent radiation. 
\item Finally, the VDE phase accompanied by halo currents that flow through surrounding structural elements.
\end{itemize}

With increasing the amount of assimilated neon in the plasma, the pre-TQ \& CQ durations as well as the IP-spike size are reduced.
While 100\%~$\textrm{D}_2$ injections show pre-TQ durations between 1--15~ms, doping the pellets with just 0.085\%~Ne already reduces the maximum duration to $< 4$~ms (with similar observations on JET~\cite{Sheikh_2024_APS}), which may pose a challenge for the staggered injection scheme for disruption mitigation at ITER.

Overall, the tendency of larger fragments (lower $\text{v}_\perp$-values) to cause shorter pre-TQ and CQ durations was reported before~\cite{Jachmich_2023_EPS, Jachmich_2024_EPS}, however, a deviating trend for the 10\%~Ne injections may hint towards a potential roll-over effect as the fragment size \& velocity are not detached in this representation.

With decreasing early CQ durations ($\Delta \textrm{t}_\textrm{CQ}^{100 \rightarrow 80}$)~\cite{Jachmich_2023_EPS, Jachmich_2024_EPS} and increasing mitigation efficiency (also of the VDE phase), this results in a continuous change in the plasma current shape during the CQ from convex (poorly or unmitigated, conduction dominated)~$\rightarrow$ concave (well mitigated, radiation dominated).

The increase of assimilated neon also affects the number of radiation peaks that can be observed via the foil bolometers.
For 100\%~$\textrm{D}_2$ injections, up to 4 smaller radiation peaks are visible which correspond to the MFA, PME, TQ, and VDE phase, respectively. As the neon content increases, the number of radiation peaks reduces until only a single, large radiation peak is visible. At the same time, the shunt current measurements also indicate reducing halo currents flowing through the lower divertor tiles. However, while for the reduced time scales only a single radiation peak is visible in the foil bolometers, we cannot conclude at this point if the other phases remain present in the chain of events.

This characterization of the SPI-induced disruption phases and their evolution with changing assimilated neon content proved as a useful basis for the interpretation of the 2022 and 2025 experiments and made comparing results across machines easier by using the visual clues -- like the CQ shape for a fast proxy of the mitigation efficiency.

\appendix
\section{IP-spike height as function of $\text{N}_\textrm{neon}$ in the pellet}

\begin{figure*}
	\centering
	\includegraphics[width=\textwidth]{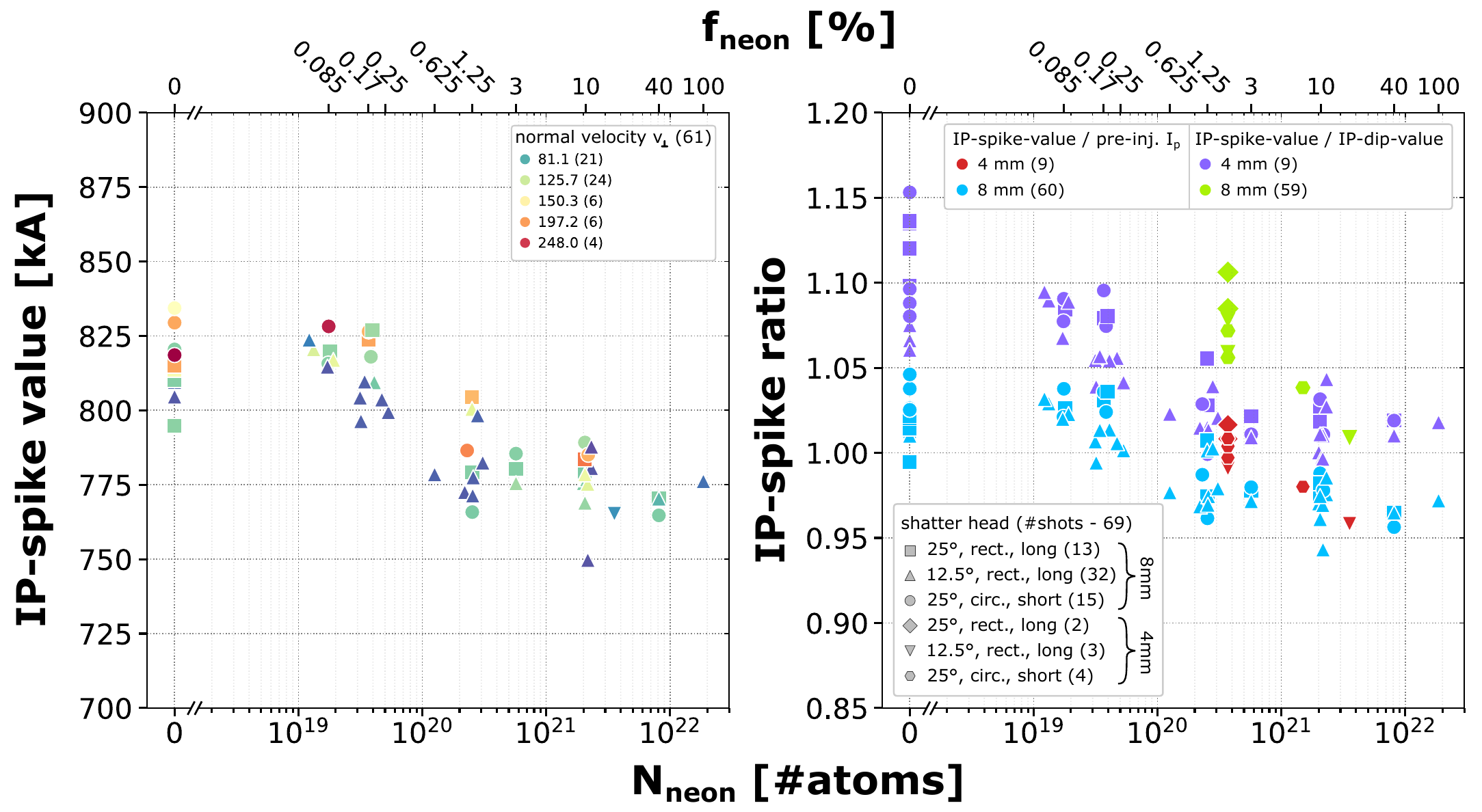}
	\caption{Scaling of the IP-spike height with neon content. The absolute (left) and relative values w.r.t. the respective pre-injection and IP-dip amplitudes (right) are provided as a function of the number of neon atoms (bottom axis) in the pellet. The top axis shows the relative pellet neon quantity. With increasing neon content, the IP-spike decreases from 105\% to 95\% of the $\approx 800$~kA pre-injection plasma current. \label{fig:IPspike}}
\end{figure*}

The IP-spike height as a function of the neon content in the pellet is shown in figure~\ref{fig:IPspike}.

\paragraph{Acknowledgements}

\noindent The authors are grateful to P.~David for his work on the bolometry system upgrade and the measurements; to A.~Bock, R.~Schramm, R.~Fischer, and B.~Kurzan for their help with the SPI experiments and V.~Igochine for fruitful discussions.
This work has been carried out within the framework of the EUROfusion Consortium, funded by the European Union via the Euratom Research and Training Programme (Grant Agreement No 101052200 — EUROfusion). Views and opinions expressed are however those of the author(s) only and do not necessarily reflect those of the European Union, the European Commission, or the ITER Organization. Neither the European Union nor the European Commission can be held responsible for them.
This work receives funding from the ITER Organization under contract IO/20/IA/43-2200. The ASDEX-Upgrade SPI project has been implemented as part of the ITER DMS Task Force programme. The SPI system and related diagnostics have received funding from the ITER Organization under contracts IO/20/CT/43-2084, IO/20/CT/43-2115, IO/20/CT/43-2116.

\bibliography{2026_Heinrich_disruption_evolution_AUG_SPI.bbl}

\end{document}